\documentclass[11pt,a4paper]{article}

\usepackage[T1]{fontenc}
\usepackage[utf8]{inputenc}
\usepackage{lmodern}
\usepackage[margin=1in]{geometry}
\usepackage{amsmath,amsthm,amssymb,mathrsfs,bm}
\usepackage{graphicx,booktabs,array,float}
\usepackage{algorithm,algorithmic}
\usepackage[authoryear,round]{natbib}
\usepackage{microtype}
\usepackage[colorlinks,citecolor=blue,linkcolor=blue,urlcolor=blue,
  bookmarksnumbered,pdfencoding=auto,psdextra]{hyperref}
\hypersetup{
  pdftitle={Robust Mixture Regressions and EM Algorithms via the Log-Truncated Likelihood},
  pdfauthor={Huiming Zhang and Longkun Hao},
  pdfsubject={Mixture regression with a log-truncated likelihood and generalized EM},
  pdfkeywords={score-induced likelihood, log-truncation, generalized EM algorithm,
    mixture regression, residual-outlier robustness}
}

\newcommand\Vector[1]{\bm{#1}}
\def\0{{\Vector{0}}}

\def\1{{\Vector{1}}}

\def\vb{{\Vector{b}}}

\def\valpha{{\Vector{\alpha}}}

\def\vpi{{\Vector{\pi}}}

\DeclareMathOperator*{\argmax}{arg\,max}

\DeclareMathOperator{\sign}{sign}

\theoremstyle{plain}
\newtheorem{theorem}{Theorem}[section]
\newtheorem{lemma}{Lemma}[section]
\newtheorem{proposition}{Proposition}[section]

\theoremstyle{definition}

\newtheorem{assumption}{Assumption}
\newtheorem{remark}{Remark}

\title{Robust Mixture Regressions and EM Algorithms via the Log-Truncated Likelihood}
\author{%
  Huiming Zhang\thanks{These authors contributed equally to this work.}\quad
  Longkun Hao\footnotemark[1]\textsuperscript{,}\thanks{Corresponding author: Longkun Hao
    (\href{mailto:sy2542103@buaa.edu.cn}{\texttt{sy2542103@buaa.edu.cn}}).}\\[0.5em]
  {\small Institute of Artificial Intelligence}\\
  {\small Beihang University, Beijing 100191, China}%
}
\date{}

\begin{document}
\maketitle

\begin{abstract}
Finite mixture regression is sensitive to gross residual outliers when component errors are modeled by Gaussian likelihoods. We propose a log-truncated likelihood for mixture regression. The likelihood is obtained by integrating a logarithmically growing Catoni-type score and exponentiating the negative integrated score, producing a proper location-scale density with a reusable normalizing constant and EM-compatible responsibilities. Although the induced density itself is lighter than exponential and has all moments, its likelihood score grows only logarithmically; this distinction yields a smooth residual-robust estimator that can be stably fitted when the data are generated from Laplace, Pareto, or even infinite-variance heavy-tailed error mechanisms. We derive deterministic tail envelopes, the location-scale decomposition of the normalizing constant, a rejection sampler, a lower-bounded mixing-weight update, a unique componentwise scale update, and a constrained generalized EM algorithm. The theory gives monotonicity of the observed likelihood and a local population-to-sample perturbation guarantee under standard compactness, contraction, and numerical-accuracy conditions. Simulations and applications to insurance claim severity and crypto return innovations show that the method is a competitive smooth likelihood-based alternative to Gaussian and pseudo-robust EM procedures, especially in residual-outlier regimes, while also clarifying its limitations under high leverage and extremely adversarial contamination.
\end{abstract}


\medskip
\noindent\textbf{MSC2020 subject classifications:} Primary 62J05, 62F35;
secondary 62H30, 62F10.

\medskip

\section{Introduction} \label{sec1}
Finite mixture regression is a standard tool for modeling heterogeneous populations
in which different latent subgroups follow different regression relationships
\citep{bai2012robust, yao2014robust, klusowski2019estimating, khalili2022overview,
naderi2023robust}. It is used in economics, insurance, finance, biology, and other
domains where a single homogeneous regression curve is inadequate
\citep{faria2013financial, paolella2015multivariate, hamel2017assessing,
mazza2020mixtures}. The classical Gaussian mixture regression model is computationally
attractive because it admits an EM algorithm with simple responsibility updates
\citep{klusowski2019estimating, kwon2020converges}, but it is also highly sensitive to
residual outliers. A small number of extreme responses can distort component
responsibilities, regression coefficients, scales, and mixing proportions.

Robust mixture regression has therefore developed along two main lines. The first
replaces Gaussian errors by more flexible likelihoods, such as
Student-$t$, Laplace, skew-normal, scale-mixture, stable, or log-concave errors
\citep{peel2000robust, yao2014robust, song2014robust, liu2014skew, zeller2016robust,
dougru2017parameter, hu2017robust, naderi2023robust, zarei2025robust}. These approaches
preserve a likelihood-based EM structure, but their performance can depend strongly on distributional tuning, initialization, and safeguards against scale degeneracy. The
second line modifies the estimating criterion through Huber, Tukey, trimmed likelihood,
or other robust losses \citep{huber1973robust, yohai1979asymptotic,
mammen1989asymptotics, neykov2007robust, bai2012robust, dougru2018robust}. These
methods can reduce sensitivity to outliers, but when inserted as pseudo-likelihood
blocks they may no longer correspond to a single normalized likelihood with coherent
responsibility, regression, and scale updates.

This paper takes a likelihood-based route. We start from the Catoni score function, a
logarithmically growing score used in robust mean estimation and log-truncated learning
\citep{catoni2012challenging, yao2022asymptotic, xu2023nonasymptotic,
zhang2025robustquasiglms}, and construct a proper error density by integrating the
score and exponentiating the negative integral. For \(1<\beta\leq2\), the score is
\[
\psi_\beta(u)=\sign(u)\log(1+|u|+|u|^\beta/\beta),
\]
and the induced density is
\[
 f(x;\alpha,\beta)=C_{\alpha,\beta}\exp\{-A_\beta(|x|/\alpha)\},~\text{with}~
 A_\beta(z)=\int_0^z\log(1+u+u^\beta/\beta)\,du.
\]
This construction keeps the likelihood structure intact. Responsibilities are computed
from a normalized density, the scale parameter has a one-dimensional likelihood update,
and the EM monotonicity argument remains available under a generalized M-step.

\paragraph{Interpretation.}
The fitted density is light-tailed; robustness comes from the score, not from tail
thickness. A central point of the paper is the separation between \emph{tail thickness
of the fitted law} and \emph{robustness of the score}. In fact,
$\log f(x;\alpha,\beta)=-\frac{\beta |x|}{\alpha}\log\frac{|x|}{\alpha}
+O(|x|/\alpha)$, so all exponential moments are finite. Robustness comes instead from the score: for
large residuals, $\psi_\beta(u)=\beta\sign(u)\log|u|+O(1)$.
Thus large residuals are downweighted relative to Gaussian likelihood, whose score
grows linearly. This makes the method suitable as a smooth likelihood-based estimator
under data generated by Laplace, Pareto, Cauchy-type or other tail-thickened
mechanisms \citep{lambert1999analysing, ahn2012new, bhati2018generalized,
zarei2025robust}.

We embed the density in a finite mixture regression model with component-specific
regression coefficients and scales. Since finite mixture likelihoods can be unbounded
when a component fits an observation exactly with vanishing scale, we work on a
constrained parameter space with lower-bounded mixing weights, bounded regression
coefficients, and lower- and upper-bounded scales. The resulting algorithm is a
constrained generalized EM algorithm. The mixing proportions are updated by a
lower-bounded simplex KKT formula. The regression coefficients are updated by projected
gradient or projected quasi-Newton steps with monotone line search. The component scales
are updated by solving a scalar equation that has a unique root whenever at least one
positively weighted residual is nonzero. An optional common-shape profile likelihood step
is included as a sensitivity analysis rather than as the primary estimator.

The contributions of this paper are as follows.
\begin{enumerate}

\item \textbf{A residual-robust generalized EM algorithm.} A logarithmically
growing Catoni-type score is converted into a proper light-tailed error density and the resulting density is used in the EM likelihood. We embed the Catoni-score-induced likelihood in finite mixture regression and derive a
generalized EM (GEM) algorithm with an explicit componentwise scale update, a
constrained mixing-weight update, and lower bounds that prevent mixture
likelihood degeneracy; see Section~\ref{sec3method}. We also give a
common-$\beta$ profile likelihood ECM step as an optional estimated-shape
extension, while retaining the fixed $\beta=2$ version as the default
robust-score baseline. The estimated-shape version is treated as a sensitivity
extension rather than as evidence about tail thickness of the fitted law.

\item \textbf{Conditional EM theory.} The theory is twofold. First, each accepted generalized M-step increases the standard
EM auxiliary function, and therefore the observed likelihood is nondecreasing. Second, under standard population contraction, compactness, and numerical error conditions \citep{balakrishnan2017statistical}, the GEM iterates remain in a local neighborhood of the correctly labeled population
solution and converge geometrically up to a statistical and numerical error radius; see
Section~\ref{sec:theory}. The result is local and conditional, matching the nonconvex nature of mixture likelihoods.
\item \textbf{Empirical evaluation.} We provide an extensive empirical evaluation on both simulated and real data.
In simulations, the proposed EM algorithm performs well under symmetric tail-thickened errors and finite-variance Pareto regimes, and profiling the common shape parameter $\beta$ can further improve performance in settings where the fixed $\beta=2$ score is too conservative.
In real-data applications, including insurance claim severity and Binance crypto
return innovations, the proposed method also delivers strong and stable empirical
performance, demonstrating its practical value; see Section~\ref{sec:real}.

\end{enumerate}
\paragraph{Organization.}
The remainder of this paper is organized as follows. Section~\ref{sec2distri} defines the Catoni-score-induced density, proves its deterministic envelopes and tail behavior, and derives the location-scale form of the normalizing constant. Section~\ref{sec3method} embeds the density in finite mixture regression and derives the constrained generalized EM algorithm, including the lower-bounded mixing-weight update, the regression block, the unique scale update, and the optional common-shape profile step. Section~\ref{sec:theory} proves the likelihood monotonicity theorem and the local sample perturbation theorem for the constrained GEM iterates. Section~\ref{sec4simulate} reports simulations under controlled residual distributions, including generalized-normal, Laplace, and Pareto regimes. Section~\ref{sec:real} presents the freMTPL claim-severity and Binance return-innovation applications.

\section{The Catoni-score-induced robust likelihood} \label{sec2distri}
In this section, we discuss the proposed score-induced density, including its formulation,
tail behavior, normalization constant, and parameter estimation.

\subsection{The Catoni-score-induced density} \label{sec21}
The inspiration for the proposed score-induced density comes from Catoni's Z-estimator
and related log-truncated robust learning methods
\citep{catoni2012challenging, chen2021generalizedcatoni, yao2022asymptotic,
zhangwei2022subweibull, xu2023nonasymptotic, zhang2025robustquasiglms}. Formally,
let \(\psi(x):\mathbb{R}\to \mathbb{R}\) be a non-decreasing anti-symmetric function
such that \(-\log(1-x+|x|^2/2) \leq \psi(x) \leq \log(1+x+|x|^2/2)\). For independent
and identically distributed (i.i.d.) samples $\{X_i\}_{i=1}^n$, \citet{catoni2012challenging}
introduced a Z-estimator for the mean satisfying \(\sum_{i=1}^{n}\psi(\frac{X_i - \mu}{\alpha}) = 0\),
where \(\alpha>0\) is a tuning parameter and \(\psi(x)\) is commonly referred to as the score
function. Taking into account the constraints imposed on the score function, a popular choice is
$\psi(x) = \sign(x)\log(1+|x| + |x|^2/2)$. Considering \(\psi(x)\) as the derivative
of \(\int_{0}^{x}\psi(t)dt\), we can similarly define Catoni's M estimate based on
\(\int_{0}^{x}\psi(t)dt\).

\begin{figure}[H]
\centering
\includegraphics[width=\textwidth]{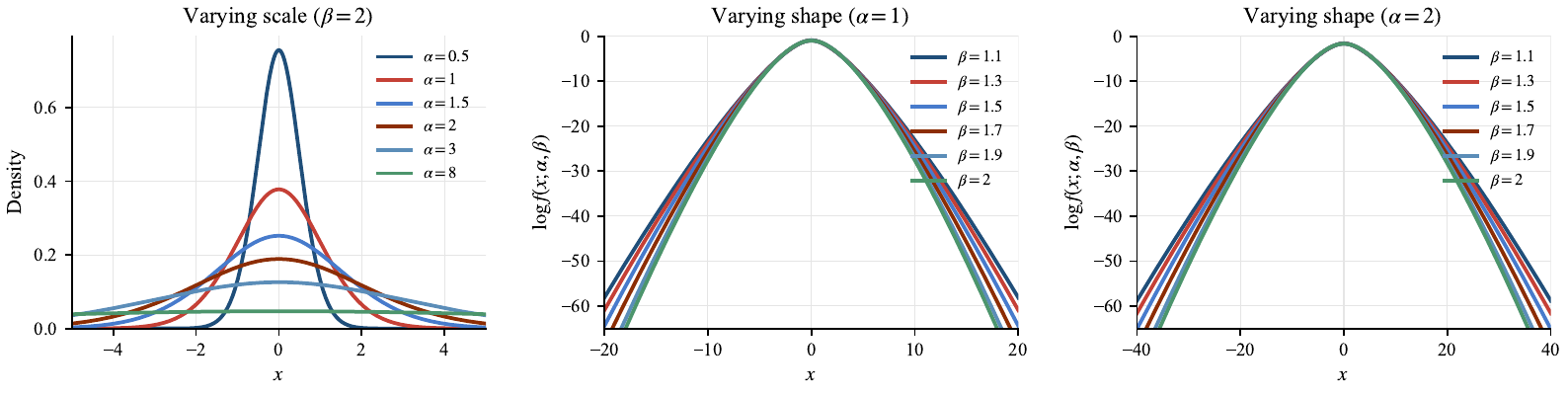}
\caption{Plots of the proposed density function under different values of
\(\alpha\) and \(\beta\). The left panel shows density curves for different
values of \(\alpha\) when \(\beta = 2\). The middle and right panels show
log-density curves for different values of \(\beta\) when \(\alpha = 1\) and
\(\alpha = 2\), respectively, making the tail-shape differences visible.}
\label{fpdffi}
\end{figure}

Consequently, \citet{chen2021generalizedcatoni} and \citet{yao2022asymptotic} derive
finite-moment and asymptotic results for Catoni-type estimators,
\citet{zhangwei2022subweibull} gives
sub-Weibull concentration and confidence-region guarantees for a log-truncated
\(Z\)-estimator without requiring a second moment, and
\citet{xu2023nonasymptotic} gives non-asymptotic guarantees for log-truncated robust
learning under infinite-variance settings. These results motivate
using a logarithmically growing score to limit the effect of extreme observations. We
begin by generalizing the score function in Catoni's Z-estimator to
\[
\psi_\beta(x)=\sign(x)\log(1+|x|+|x|^\beta/\beta),
\qquad 1<\beta\leq2 .
\]
For \(z\geq0\), define
\[
A_\beta(z)=\int_0^z\log(1+u+u^\beta/\beta)\,du,
\qquad R_\beta(x)=2A_\beta(|x|).
\]
Equivalently, \(R_\beta(x)=2\int_0^x\psi_\beta(t)\,dt\), since
\(\psi_\beta\) is odd. The absolute-value definition makes the symmetry of the loss
explicit. For a scale parameter \(\alpha>0\), the proposed density is
\begin{equation}\label{eq:proposed-density}
    f(x;\alpha,\beta)
=C_{\alpha,\beta}\exp\{-A_\beta(|x|/\alpha)\}
=C_{\alpha,\beta}\exp\{-R_\beta(x/\alpha)/2\},
\end{equation}
where
\(C^{-1}_{\alpha,\beta}=\int_{-\infty}^{\infty}
\exp\{-A_\beta(|t|/\alpha)\}\,dt\).
 Figure \ref{fpdffi} shows the plot of \(f(x; \alpha, \beta)\) under different scale parameter \(\alpha\) and shape parameter \(\beta\).
Sampling from the Catoni-score-induced density \eqref{eq:proposed-density} is presented in \ref{app:sampling} by an envelope-based rejection sampler.

The first property records the tail behavior of the density. As $|x|\to\infty$, the log density is of order
$$\log f(x;\alpha,\beta)=-\bigl(1+o(1)\bigr)\frac{\beta|x|}{\alpha}\log\frac{|x|}{\alpha},$$
so the proposed fitted law is lighter than every
exponential distribution; see \ref{app:tail_behavior} for details. Catoni-type
M-estimators can nevertheless reduce the effect of extreme residuals because their
robustness comes from the logarithmically growing likelihood score.

\begin{lemma}\label{upbound}
Let \(\phi(z)=(1+z)\log(1+z)-z\) for \(z\geq0\). For any
\(\alpha>0, \beta\in (1, 2]\),
\[
2\phi(|x|/\alpha)
\leq R_\beta(x/\alpha)
\leq \frac{x^2}{\alpha^2}
+\frac{2|x|^{\beta+1}}{\beta(\beta+1)\alpha^{\beta+1}},~x \in \mathbb{R}.
\]
\end{lemma}
Lemma \ref{upbound} gives deterministic envelopes for the integrated score; its proof is
given in~\ref{app:proof_upbound}. The lower
bound provides the rejection-sampling envelope in~\ref{app:sampling}, while
the upper bound is useful for optimization and numerical arguments. Neither is the exact tail asymptotic of the density; the tail behavior is given by Proposition~\ref{prop:tail_behavior}. For computational reasons, we next discuss the
calculation of the normalization constant and parameter estimation of \eqref{eq:proposed-density}.

The normalization constant plays an important role in likelihood evaluation, parameter
estimation, and sampling from the proposed distribution. Although it does not admit a
closed-form expression, its dependence on the scale parameter admits a simple location-scale decomposition. Recall that
\(C^{-1}_{\alpha,\beta}
=\int_{-\infty}^{\infty}\exp\{-R_\beta(x/\alpha)/2\}\,dx\).
Since $\psi_\beta$ is an odd function, $R_\beta(x)=2\int_0^x\psi_\beta(t)dt$ is an even
function. Therefore, by the change of variables $z=x/\alpha$, we prove
\begin{equation}\label{eq:C_alpha_beta}
C_{\alpha,\beta}
=
\frac{1}{\alpha I_\beta},
\end{equation}
where \(I_\beta\) is the scale-one
normalizing integral derived in~\ref{app:norm_const}. \eqref{eq:C_alpha_beta} shows that, for a fixed shape parameter $\beta$, the
normalization constant only needs to be computed once up to the scale factor $\alpha$.
In particular, when $\beta$ is fixed, $I_\beta$ can be
precomputed numerically and reused by the EM iterations; see \ref{app:norm_const} for details.

\subsection{Parameter estimation}
\label{sec:dist_param_est}

For completeness, we briefly discuss how to estimate the parameters of the proposed
distribution from independent observations $\{X_i\}_{i=1}^n\overset{\text{i.i.d.}}{\sim} f(\cdot;\alpha,\beta)$.
The likelihood is 
\begin{equation}
   L(\alpha,\beta)
=
\prod_{i=1}^n
C_{\alpha,\beta}
\exp\left\{
-\frac{1}{2}R_\beta(x_i/\alpha)
\right\}.
\end{equation}

Using the scale decomposition $C_{\alpha,\beta}=1/(\alpha I_\beta)$, the log-likelihood can be written as
\(\ell(\alpha,\beta)=-n\log\alpha-n\log I_\beta
-\frac{1}{2}\sum_{i=1}^nR_\beta(x_i/\alpha)\).
This representation separates the scale parameter $\alpha$ from the shape-dependent
normalizing factor $I_\beta$. Therefore, for a fixed $\beta$, estimating $\alpha$ reduces
to a one-dimensional numerical optimization problem. When $\beta$ is also estimated, a
stable approach is to profile out $\alpha$: for each candidate
$\beta\in[\beta_{\min},2]$, compute the scale maximizer and then maximize the profiled
likelihood over the one-dimensional shape interval. We use this profile likelihood idea
as the estimated-$\beta$ extension in Section~\ref{sec3method}. Because $\beta$ can be
weakly identified after profiling scale, the fixed $\beta=2$ model remains the default
baseline and estimated $\beta$ is reported as an ablation or diagnostic.
The derivative formula used by the scale update is provided in ~\ref{app:alpha_update}.

\section{GEM algorithm}
\label{sec3method}

We now apply the proposed robust score-induced density to mixture regression. Let
$\{(X_j,Y_j)\}_{j=1}^n$ be independent observations with $X_j\in\mathbb{R}^p$ and
$Y_j\in\mathbb{R}$. Suppose that the data are generated from an $m$-component mixture
regression model
\begin{equation}\label{mixture}
Y_j=X_j^\top b_i+\epsilon_{ij},\qquad \epsilon_{ij}\sim f(\cdot;\alpha_i,\beta),
\quad \text{with probability } \pi_i ,
\end{equation}
where $\pi_i>0$, $\sum_{i=1}^m\pi_i=1$, $b_i\in\mathbb{R}^p$, and $\alpha_i>0$.
The default implementation fixes $\beta=2$, following the Catoni-type score that
motivates the proposed density. We also give an optional common-$\beta$ update below;
component-specific $\beta_i$ updates are not used because small mixture components give
unstable effective sample sizes for shape estimation.

Finite mixture likelihoods with component-specific scales can be unbounded if a component
fits one observation exactly and its scale tends to zero. We therefore state the algorithm
on the constrained parameter space
\[
\Omega=
\left\{
\Theta:\ \pi_i\geq \pi_{\min},\ \sum_{i=1}^m\pi_i=1,\ \|b_i\|_2\leq B,\ 
\alpha_{\min}\leq\alpha_i\leq\alpha_{\max},\ i=1,\ldots,m
\right\},
\]
where $0<\pi_{\min}<1/m$, $0<\alpha_{\min}<\alpha_{\max}<\infty$, and $B<\infty$ are
fixed constants. The lower scale bound is a standard safeguard against the scale
degeneracy of mixture likelihoods. If $\beta$ is estimated, we use the extended compact
space $\Omega_\beta=\Omega\times[\beta_{\min},2]$ with $\beta_{\min}>1$.

\subsection{Latent allocation and structural updates}
Let $\Theta=(\pi_1,\ldots,\pi_m,b_1,\ldots,b_m,\alpha_1,\ldots,\alpha_m)$ for fixed
$\beta$; in the estimated-shape extension, $\beta$ is appended to $\Theta$. The observed
log-likelihood is
\begin{equation}\label{loglikelihood}
\ell(Y,X\mid\Theta)
=
\sum_{j=1}^n
\log\left\{
\sum_{i=1}^m
\pi_i f(Y_j-X_j^\top b_i;\alpha_i,\beta)
\right\}.
\end{equation}
We estimate $\Theta$ by an EM-type algorithm. Given the current estimate
$\Theta^{(t)}$, the E-step computes the posterior responsibility
\begin{equation}\label{eq:responsibility}
w_{ij}^{(t)}
=
\frac{
\pi_i^{(t)}f(Y_j-X_j^\top b_i^{(t)};\alpha_i^{(t)},\beta)
}{
\sum_{\ell=1}^m
\pi_\ell^{(t)}f(Y_j-X_j^\top b_\ell^{(t)};\alpha_\ell^{(t)},\beta)
},
\qquad i=1,\ldots,m,\ j=1,\ldots,n .
\end{equation}
The corresponding auxiliary function is
\begin{equation}\label{eq:Q_function}
Q(\Theta\mid\Theta^{(t)})
=
\sum_{j=1}^n\sum_{i=1}^m
w_{ij}^{(t)}
\left[
\log\pi_i-\log w_{ij}^{(t)}
+\log C_{\alpha_i,\beta}
-\frac12
R_\beta\left(\frac{Y_j-X_j^\top b_i}{\alpha_i}\right)
\right].
\end{equation}

Since the joint maximization of \eqref{eq:Q_function} has no closed-form solution, we
perform a blockwise generalized M-step over $\pi$, $b$, $\alpha$, and optionally
$\beta$. The mixing proportions are updated by the constrained simplex problem
\begin{equation}\label{pistep}
\max_{\pi_i\geq\pi_{\min},\ \sum_i\pi_i=1}
\sum_{i=1}^m W_i^{(t)}\log\pi_i,
\qquad
W_i^{(t)}=\sum_{j=1}^n w_{ij}^{(t)} .
\end{equation}
The solution is the lower-bounded simplex KKT update
\begin{equation}\label{eq:pi_kkt_update}
\pi_i^{(t+1)}
=
\max\left\{\pi_{\min},\frac{W_i^{(t)}}{\lambda_t}\right\},
\qquad
\sum_{i=1}^m
\max\left\{\pi_{\min},\frac{W_i^{(t)}}{\lambda_t}\right\}=1 ,
\end{equation}
where $\lambda_t>0$ is found by a one-dimensional bisection search. If no lower-bound
constraint is active, then $\lambda_t=n$ and \eqref{eq:pi_kkt_update} reduces to the
usual EM update $W_i^{(t)}/n$. The following proposition gives the exact
lower-bounded update for the mixing proportions; its proof is given in
\ref{app:proof_pi_step}.
\begin{proposition}[Lower-bounded mixing-weight update]\label{prop:pi_lower_bound}
Assume $0<\pi_{\min}<1/m$ and $W_i^{(t)}\geq0$ with $\sum_iW_i^{(t)}=n$. The optimizer of
\eqref{pistep} is given by \eqref{eq:pi_kkt_update}, with the unique $\lambda_t>0$
satisfying the simplex constraint.
\end{proposition}
Operationally, $\pi_{\min}$ acts as a floor on the effective size of each fitted
component. When $W_i^{(t)}$ is very small, the projected update keeps
$\pi_i^{(t+1)}$ at $\pi_{\min}$ instead of allowing the component to disappear. This is
important because a vanished component receives negligible responsibilities in later
iterations, so its regression and scale parameters are no longer updated from a stable
effective sample. The lower-bounded simplex update therefore improves numerical
stability while preserving the compactness assumptions used in the generalized EM
convergence argument.

Given $\pi^{(t+1)}$ and $\alpha^{(t)}$, the regression coefficients are updated by
minimizing
\begin{equation}\label{bstep}
\min_{\|b_i\|_2\leq B,\ i=1,\ldots,m}
\sum_{j=1}^n\sum_{i=1}^m
\frac12 w_{ij}^{(t)}
R_\beta\left(\frac{Y_j-X_j^\top b_i}{\alpha_i^{(t)}}\right).
\end{equation}
We solve \eqref{bstep} by projected gradient or projected quasi-Newton iterations with a
monotone line search. Each trial step is projected onto the Euclidean ball
$\{b_i:\|b_i\|_2\leq B\}$ and accepted only if it does not decrease the auxiliary
function; the inner iteration stops when the projected-gradient norm is below
$\delta_b$ or when a fixed maximum number of inner iterations is reached. The gradient
with respect to $b_i$ before projection is
\begin{equation}\label{eq:b_gradient}
\nabla_{b_i}
=
-\sum_{j=1}^n
w_{ij}^{(t)}
\frac{X_j}{\alpha_i^{(t)}}
\psi_\beta\left(\frac{Y_j-X_j^\top b_i}{\alpha_i^{(t)}}\right).
\end{equation}
\subsection{Robustness scope of the estimator}
\label{subsec:influence_scope}

The logarithmic score reduces, but does not bound, the effect of an extreme residual.
This distinction can be made precise in the single-component fixed-scale case. Let
\(b_0\) solve
\[
\Psi(b;P)
=
\mathbb E_P\left[
\frac{X}{\alpha}
\psi_\beta\left(\frac{Y-X^\top b}{\alpha}\right)
\right]=0
\]
and suppose
$M
=
\mathbb E_P\left[
\frac{XX^\top}{\alpha^2}
\psi_\beta'\left(\frac{Y-X^\top b_0}{\alpha}\right)
\right]$ is nonsingular. Standard differentiation of the estimating equation under point-mass
contamination at \((x,y)\) gives the influence function
\begin{equation}\label{eq:influence_single_component}
\operatorname{IF}\{(x,y);b,P\}
=
M^{-1}\frac{x}{\alpha}
\psi_\beta\left(\frac{y-x^\top b_0}{\alpha}\right).
\end{equation}
As \(|u|\to\infty\),
\(\psi_\beta(u)=\beta\sign(u)\log|u|+O(1)\). Hence
\eqref{eq:influence_single_component} remains unbounded as the residual diverges,
although it grows only logarithmically rather than linearly as under Gaussian
likelihood. It is also unbounded as \(\|x\|_2\) grows. 

\subsection{Componentwise scale and shape updates}
\label{subsec:scale_shape}

Given $\pi^{(t+1)}$ and $b^{(t+1)}$, the scale parameters are updated component
by component. Let $r_{ij}^{(t+1)}=Y_j-X_j^\top b_i^{(t+1)}$. For component $i$, the
scale objective is
\begin{equation}\label{eq:alpha_component_objective}
Q_i^{(t)}(\alpha_i)
=
\sum_{j=1}^n
w_{ij}^{(t)}
\left[
\log C_{\alpha_i,\beta}
-\frac12
R_\beta\left(\frac{r_{ij}^{(t+1)}}{\alpha_i}\right)
\right],
\qquad \alpha_i>0 .
\end{equation}
Using $C_{\alpha,\beta}=1/(\alpha I_\beta)$, the first-order condition for
\eqref{eq:alpha_component_objective} can be written as \(G_i^{(t)}(\alpha_i)=0\), where
\(G_i^{(t)}\) is defined below. The
following theorem gives the scale update used by the constrained GEM algorithm.

\begin{theorem}[Constrained scale update]\label{thm:alpha_root_componentwise}
Fix $\beta\in(1,2]$, suppose
$W_i^{(t)}=\sum_{j=1}^n w_{ij}^{(t)}>0$ for component $i$ and define
\begin{equation}\label{eq:alpha_root}
G_i^{(t)}(\alpha_i)
:=
\sum_{j=1}^n
w_{ij}^{(t)}
\frac{|r_{ij}^{(t+1)}|}{\alpha_i}
\log\!\left(
1+\frac{|r_{ij}^{(t+1)}|}{\alpha_i}
+\frac{1}{\beta}
\left(\frac{|r_{ij}^{(t+1)}|}{\alpha_i}\right)^\beta
\right)
-
W_i^{(t)} .
\end{equation}
If $\sum_{j=1}^n w_{ij}^{(t)}
\mathbf{1}\{|r_{ij}^{(t+1)}|>0\}>0,$
 then, 
\begin{center}
$G_i^{(t)}$ is continuous and strictly decreasing on $(0,\infty)$ and has a unique
root $\widetilde\alpha_i\in(0,\infty)$,
\end{center}
which is the unique unconstrained maximizer of
$Q_i^{(t)}(\alpha_i)$. On the constrained interval
$[\alpha_{\min},\alpha_{\max}]$, the maximizer is the clipped value of
\(\widetilde\alpha_i\) in this interval.
If all positively weighted residuals are zero, then $Q_i^{(t)}(\alpha_i)$ has no
unconstrained maximizer and the constrained maximizer is $\alpha_i^{(t+1)}=\alpha_{\min}$.   
\end{theorem}

Theorem~\ref{thm:alpha_root_componentwise} reduces the $\alpha_i$-step to a
one-dimensional root-finding problem followed by projection onto the admissible scale
interval. In practice, we solve \eqref{eq:alpha_root} by bisection or by Brent's
safeguarded root-finding method \citep{brent1973algorithms}, with an initial bracket
obtained from the weighted MAD of the residuals and the fixed limits
$[\alpha_{\min},\alpha_{\max}]$. Here, for component \(i\),
\(\operatorname{Med}_{w_i}(\cdot)\) denotes the weighted \(0.5\)-quantile using the
positive weights \(w_{ij}^{(t)}\),
\[
m_i=\operatorname{Med}_{w_i}\{r_{ij}^{(t+1)}\},\qquad
\operatorname{MAD}_{w_i}(r_i)
=
\frac{
\operatorname{Med}_{w_i}\{|r_{ij}^{(t+1)}-m_i|\}
}{\Phi^{-1}(0.75)}.
\]
The denominator \(\Phi^{-1}(0.75)\approx0.6745\) makes the MAD scale equal to the
standard deviation under a Gaussian reference distribution.
Influence-function-based concentration analyses of robust M-estimators motivate
controlled tuning in finite samples \citep{mathieu2022concentration}. In our
constrained GEM formulation, the fixed admissible scale range additionally keeps the
componentwise root search well posed and numerically stable. The proof is given in
\ref{app:proof_alpha_root}.

When the shape parameter is estimated, we use a common-$\beta$ profile likelihood block
rather than separate component-specific shapes. 
For fixed responsibilities and updated residuals $r_{ij}^{(t+1)}$, let
$\widehat\alpha_i(\beta)$ be the constrained scale maximizer obtained from
Theorem~\ref{thm:alpha_root_componentwise} with this candidate $\beta$. The profiled
shape objective function is
\begin{equation}\label{eq:beta_profile_objective_main}
Q_p^{(t)}(\beta)
=
\sum_{i=1}^m\sum_{j=1}^n
w_{ij}^{(t)}
\left[
-\log\widehat\alpha_i(\beta)
-\log I_\beta
-A_\beta\left(\frac{|r_{ij}^{(t+1)}|}{\widehat\alpha_i(\beta)}\right)
\right],~
\beta\in[\beta_{\min},2].
\end{equation}
The estimated-shape update chooses
\begin{equation}\label{eq:beta_profile_update}
\beta_{\mathrm{cand}}^{(t+1)}
\in
\argmax_{\beta\in[\beta_{\min},2]} Q_p^{(t)}(\beta),
\end{equation}
using a grid search initialized by a robust trimmed-moment, with
optional Brent or damped Newton refinement inside the bracket \citep{brent1973algorithms,nocedal2006numerical}. The candidate pair
$\{\widehat\alpha_i(\beta_{\mathrm{cand}}^{(t+1)})\}_{i=1}^m$ and
$\beta_{\mathrm{cand}}^{(t+1)}$ is accepted only if the EM auxiliary objective does not
decrease; otherwise the algorithm retains the previous $\beta$ and performs the
fixed-$\beta$ scale update. This profile ECM step is the likelihood-efficient target
under the correctly specified density, while the robust pilot and monotone acceptance are
used to guard against finite-sample weak identification and boundary solutions. The
derivative of this optional profile block is recorded in~\ref{app:beta_profile_derivative}; because lower-bound selections indicate
weak identification and \eqref{eq:beta_profile_update} is also treated as
a sensitivity analysis in the experiments.

\begin{algorithm}[h]
\caption{Generalized EM algorithm for the constrained mixture model \eqref{mixture}.}
\label{EMrobust}
\small
\begin{algorithmic}[1]
\REQUIRE Data $\{(Y_j,X_j)\}_{j=1}^n$, number of components $m$, fixed $\beta$ or initial $\beta^{(0)}\in[\beta_{\min},2]$, constraints $\pi_{\min},B,\alpha_{\min},\alpha_{\max}$, tolerance $\delta$, and initial value $\Theta^{(0)}\in\Omega$.
\ENSURE Estimated parameter $\hat{\Theta}$.
\STATE Compute $\ell(\Theta^{(0)})$; set $t=0$. If $\beta$ is fixed, set $\beta^{(t)}\equiv\beta$.
\REPEAT
\STATE \textbf{E-step:} Compute responsibilities \(w_{ij}^{(t)}\) by \eqref{eq:responsibility}, with \(\beta\) replaced by \(\beta^{(t)}\) if it is estimated.
\STATE \textbf{$\vpi$-step:} Let \(W_i^{(t)}=\sum_{j=1}^n w_{ij}^{(t)}\). Find \(\lambda_t>0\) satisfying \(\sum_i\max\{\pi_{\min},W_i^{(t)}/\lambda_t\}=1\), and set \(\pi_i^{(t+1)}=\max\{\pi_{\min},W_i^{(t)}/\lambda_t\}\).
\STATE \textbf{$\vb$-step:} Update \(\vb^{(t+1)}\) by projected gradient or projected quasi-Newton iterations for \eqref{bstep}, accepting only line-search steps that do not decrease \(Q\).
\STATE \textbf{$\valpha/\beta$-step:} If \(\beta\) is fixed, solve \(G_i^{(t)}(\alpha_i)=0\) for each component when possible and clip to \([\alpha_{\min},\alpha_{\max}]\) as in Theorem~\ref{thm:alpha_root_componentwise}. If \(\beta\) is estimated, compute the profile candidate \eqref{eq:beta_profile_update}, update the corresponding \(\valpha\), and accept the pair only when the auxiliary objective does not decrease.
\STATE Compute $\ell(\Theta^{(t+1)})$.
\STATE Set $t\leftarrow t+1$.
\UNTIL{$|\ell(\Theta^{(t)})-\ell(\Theta^{(t-1)})|\le \delta$}
\STATE \textbf{Output:} $\hat{\Theta}=\Theta^{(t)}$.
\end{algorithmic}
\end{algorithm}

Although the M-step is implemented blockwise, each block update is chosen so as not to
decrease the auxiliary function. The $\vpi$-step is solved exactly by
Proposition~\ref{prop:pi_lower_bound}, the $\vb$-step is performed with projection and a
monotone line search, and the $\valpha$-step is solved as the unique componentwise
maximizer given by Theorem~\ref{thm:alpha_root_componentwise}. When the optional
common-$\beta$ profile block is used, it is accepted only after verifying that the same
auxiliary function is not decreased. Hence the
blockwise M-step is a generalized EM update:
\begin{equation}\label{eq:GEM_condition}
Q(\Theta^{(t+1)}\mid\Theta^{(t)})
\geq
Q(\Theta^{(t)}\mid\Theta^{(t)}).
\end{equation}

\section{Convergence theory for the Generalized EM algorithm}\label{sec:theory}
The previous section turns the log-truncated-score-induced likelihood into a constrained
blockwise GEM algorithm. We now record the theoretical guarantees that justify this
construction. The first result shows that the accepted block updates preserve the basic
EM monotonicity property, while the second result states a local finite-sample
perturbation bound under standard population contraction and uniform-deviation
conditions. These results are deliberately conditional: they rule out scale degeneracy
through the compact constraints and address local behavior near a correctly labelled
population solution, rather than claiming global optimization of the non-convex mixture
likelihood.

\begin{theorem}[Monotonicity of the generalized EM algorithm]
\label{EMmonotonically1}
If the update at iteration $t$ satisfies \eqref{eq:GEM_condition}, then the observed-data
log-likelihood is non-decreasing:
\begin{equation}
   \ell(Y,X\mid\Theta^{(t+1)})\geq\ell(Y,X\mid\Theta^{(t)}).
\end{equation}

\end{theorem}
The proof is given in~\ref{app:proof_em_monotonicity}.

The monotonicity result alone does not imply convergence of the parameter sequence. We
therefore state separate local population and sample perturbation guarantees for the
constrained algorithm. These results follow the population-to-sample EM framework of
\citet{balakrishnan2017statistical}; they are local and conditional, rather than global optimality guaranties for the non-convex mixture likelihood.

For the fixed-\(\beta\) algorithm, work in a correctly labelled local chart and write
\[
\theta=
(\pi_1,\ldots,\pi_{m-1}, b_1^\top,\ldots,b_m^\top,\alpha_1,\ldots,\alpha_m)^\top
\in\mathbb{R}^{d_\theta},
\qquad d_\theta=m(p+2)-1,
\]
where \(\pi_m=1-\sum_{i=1}^{m-1}\pi_i\). If the optional common-shape profile step is analyzed,
the chart appends the common coordinate \(\beta\in[\beta_{\min},2]\), and the same
assumptions must hold in that extended chart. Let \(\theta^\star\) be the correctly
labelled population parameter and define
\[
\mathbb B(r;\theta^\star)=
\{\theta:\|\theta-\theta^\star\|_2\le r\}.
\]
 All
maximizations below are over this same local chart.
For \(O=(Y,X)\) and \(\theta\in{\mathbb B(r;\theta^\star)}\), define the population
responsibility
\[
w_i(O;\theta)
=
\frac{\pi_i f(Y-X^\top b_i;\alpha_i,\beta)}
{\sum_{\ell=1}^{m}\pi_\ell f(Y-X^\top b_\ell;\alpha_\ell,\beta)} .
\]
To avoid ambiguity, we distinguish the full Jensen auxiliary function from the reduced
objective used to define the M-step map. For a candidate new parameter \(\theta'\) and
old parameter \(\theta\), the full population auxiliary function is
\[
Q^{f}(\theta'\mid\theta)
=
\mathbb E_{\theta^\star}
\sum_{i=1}^m
w_i(O;\theta)
\Bigl\{
\log \pi_i'
+\log f(Y-X^\top b_i';\alpha_i',\beta)
-\log w_i(O;\theta)
\Bigr\},
\]
and its sample version is
\[
Q_n^{{f}}(\theta'\mid\theta)
=
\frac1n\sum_{j=1}^n\sum_{i=1}^m
w_i(O_j;\theta)
\Bigl\{
\log \pi_i'
+\log f(Y_j-X_j^\top b_i';\alpha_i',\beta)
-\log w_i(O_j;\theta)
\Bigr\}.
\]
The last term in each display is the conditional entropy term. Define
\[
H(\theta)
=
-\mathbb E_{\theta^\star}
\sum_{i=1}^m w_i(O;\theta)\log w_i(O;\theta),
\qquad
H_n(\theta)
=
-\frac1n\sum_{j=1}^n\sum_{i=1}^m
w_i(O_j;\theta)\log w_i(O_j;\theta).
\]
Removing this entropy term gives the reduced population and sample objectives
\begin{align}
\bar Q(\theta'\mid \theta)
&=
\mathbb{E}_{\theta^\star}
\sum_{i=1}^{m}
w_i(O;\theta)
\Bigl\{
\log\pi_i'
+
\log f(Y-X^\top b_i';\alpha_i',\beta)
\Bigr\},\label{eq:population_Q_robust}\\
\bar Q_n(\theta'\mid \theta)
&=
\frac1n\sum_{j=1}^{n}
\sum_{i=1}^{m}
w_i(O_j;\theta)
\Bigl\{
\log\pi_i'
+
\log f(Y_j-X_j^\top b_i';\alpha_i',\beta)
\Bigr\}.\label{eq:sample_Q_robust}
\end{align}
Thus the full and reduced objectives satisfy
\[
Q^{f}(\theta'\mid\theta)
=
\bar Q(\theta'\mid\theta)+H(\theta),
\qquad
Q_n^{f}(\theta'\mid\theta)
=
\bar Q_n(\theta'\mid\theta)+H_n(\theta).
\]

For fixed parameter \(\theta\), \(H(\theta)\) and \(H_n(\theta)\) depend only on $\theta$, not on
\(\theta'\). Hence
\[
\argmax_{\theta'\in{\mathbb B(r;\theta^\star)}} Q^{f}(\theta'\mid\theta)
=
\argmax_{\theta'\in{\mathbb B(r;\theta^\star)}} \bar Q(\theta'\mid\theta),~~
\argmax_{\theta'\in{\mathbb B(r;\theta^\star)}} Q_n^{f}(\theta'\mid\theta)
=
\argmax_{\theta'\in{\mathbb B(r;\theta^\star)}} \bar Q_n(\theta'\mid\theta).
\]
The local population EM operator and the exact local sample EM operator are therefore
defined explicitly by
\[
M(\theta)\in
\argmax_{\theta'\in{\mathbb B(r;\theta^\star)}}\bar Q(\theta'\mid\theta),
\qquad
M_n(\theta)\in
\argmax_{\theta'\in{\mathbb B(r;\theta^\star)}}\bar Q_n(\theta'\mid\theta).
\]
Equivalently, the same operators are obtained by maximizing
\(Q^{f}\) and \(Q_n^{f}\), respectively, because the entropy
terms are constants with respect to the M-step argument \(\theta'\). In the fixed-\(\beta\)
algorithm, \(\beta\) is not part of \(\theta'\); in the optional common-shape chart, the
candidate density in the displays above is read with the candidate common shape
\(\beta'\), while the responsibilities remain those computed from the old parameter
\(\theta\). Write \(\mathcal A_n\) for one execution of the E-step and accepted
blockwise generalized M-step in Algorithm~\ref{EMrobust}. Starting from \(\theta\), the
implemented update is
\begin{equation}\label{eq:implemented_sample_operator}
\widetilde M_n(\theta)
:=
\mathcal A_n(\theta)
=
\theta^+,
\qquad
\theta^+\in{\mathbb B(r;\theta^\star)},\qquad
\bar Q_n(\theta^+\mid\theta)\ge \bar Q_n(\theta\mid\theta).
\end{equation}
The last inequality is the GEM acceptance condition. Thus \(\widetilde M_n=M_n\) when
the M-step is solved exactly over \({\mathbb B(r;\theta^\star)}\); otherwise
\(\|\widetilde M_n(\theta)-M_n(\theta)\|_2\) measures the numerical error of the
implemented generalized M-step.

\begin{assumption}[Local identifiability, compact chart, and curvature]
\label{ass:local_chart_curvature}
There exist constants \(r>0\), \(0<\pi_{\min}<1/m\),
\(B<\infty\), and scale bounds \(\alpha_{\min},\alpha_{\max}\) such that, in the convex local chart
containing \(\theta^\star\), all parameters satisfy
\(\pi_i\ge \pi_{\min}\), \(\|b_i\|_2\le B\), and
\(0<\alpha_{\min}\le \alpha_i\le \alpha_{\max}<\infty\). The local labeling is identifiable
in \({\mathbb B(r;\theta^\star)}\). Moreover, with
\(q(\theta')=\bar Q(\theta'\mid\theta^\star)\), the map \(q\) is
\(\lambda\)-strongly concave on \({\mathbb B(r;\theta^\star)}\), and the first-order
stability condition holds: there is \(\gamma\in[0,\lambda)\) such that
\begin{equation}\label{eq:FOS_robust}
\left\|
\nabla_{\theta'}\bar Q(M(\theta)\mid\theta^\star)
-
\nabla_{\theta'}\bar Q(M(\theta)\mid\theta)
\right\|_2
\le
\gamma\|\theta-\theta^\star\|_2,
\qquad \theta\in{\mathbb B(r;\theta^\star)}.
\end{equation}
\end{assumption}

\begin{theorem}[Population contraction]
\label{thm:robust_population_contraction}
Under Assumption~\ref{ass:local_chart_curvature}, the population EM operator is locally
contractive. More precisely, with \(\kappa=\gamma/\lambda<1\),
\begin{equation}\label{eq:population_contraction_robust}
\|M(\theta)-\theta^\star\|_2
\le
\kappa\|\theta-\theta^\star\|_2,
\qquad
\theta\in{\mathbb B(r;\theta^\star)}.
\end{equation}
Consequently, the population iterates \(\theta^{t+1}=M(\theta^t)\) satisfy
\[
\|\theta^t-\theta^\star\|_2
\le
\kappa^t\|\theta^0-\theta^\star\|_2,
\qquad t=1,2,\ldots .
\]
\end{theorem}

For \(\delta\in(0,1)\), define the uniform sample deviation of the local EM operator by
\begin{equation}\label{eq:epsilon_unif_robust}
\epsilon_M^{\mathrm{unif}}(n,\delta)
=
\inf\left\{
\epsilon>0:
\mathbb P\left(
\sup_{\theta\in{\mathbb B(r;\theta^\star)}}
\|M_n(\theta)-M(\theta)\|_2\le \epsilon
\right)\ge 1-\delta
\right\}.
\end{equation}
For the implemented generalized M-step, define the deterministic or high-probability
inner optimization error
\begin{equation}\label{eq:algorithmic_error_robust}
a_n
=
\sup_{\theta\in{\mathbb B(r;\theta^\star)}}
\|\widetilde M_n(\theta)-M_n(\theta)\|_2 .
\end{equation}
Thus \(a_n\) is the worst-case distance, over the local chart, between the exact sample
EM update and the blockwise numerical update actually returned by the algorithm. If the
\(b\)-step and the \(\alpha\)-step are solved to numerical precision, \(a_n\) can be made
as small as the prescribed inner tolerance; in the ideal exact-block case \(a_n=0\).

\begin{theorem}[Sample-based convergence of Algorithm~\ref{EMrobust}]
\label{thm:robust_sample_em}
Suppose that Assumption~\ref{ass:local_chart_curvature} holds and let
\(\kappa=\gamma/\lambda\). Assume that \(\theta^0\in{\mathbb B(r;\theta^\star)}\) and that
\begin{equation}\label{eq:sample_condition_robust}
\epsilon_M^{\mathrm{unif}}(n,\delta)+a_n
\le
(1-\kappa)r .
\end{equation}
Then the iterates generated by the implemented update
\(\theta^{t+1}=\widetilde M_n(\theta^t)\) satisfy, with probability at least
\(1-\delta\),
\begin{equation}\label{eq:sample_bound_robust}
\|\theta^t-\theta^\star\|_2
\le
\kappa^t\|\theta^0-\theta^\star\|_2
+
\frac{\epsilon_M^{\mathrm{unif}}(n,\delta)+a_n}{1-\kappa},
\qquad t=1,2,\ldots .
\end{equation}
In particular, after
$
T\ge
\log_{1/\kappa}
\left(
\frac{(1-\kappa)\|\theta^0-\theta^\star\|_2}
{\epsilon_M^{\mathrm{unif}}(n,\delta)+a_n}
\right)
$
iterations, the estimator satisfies
\begin{equation}\label{eq:sample_terminal_robust}
\|\theta^T-\theta^\star\|_2
\le
\frac{2\{\epsilon_M^{\mathrm{unif}}(n,\delta)+a_n\}}{1-\kappa}.
\end{equation}
\end{theorem}

The next theorem turns the abstract deviation term into a finite-dimensional statistical
radius for the proposed robust mixture regression.
\begin{theorem}[A non-asymptotic robust mixture regression rate]
\label{thm:robust_rate}
Assume, in addition to Assumption~\ref{ass:local_chart_curvature}, that \(X\) is
sub-Gaussian, that the model \eqref{mixture} is correctly specified at \(\theta^\star\),
and that the local Hessian
\(-\nabla^2_{\theta'\theta'}\bar Q(\theta'\mid\theta)\) is uniformly bounded below by
\(\lambda_0 I\) on
\({\mathbb B(r;\theta^\star)}\times{\mathbb B(r;\theta^\star)}\) for some
\(\lambda_0>0\). Then there are constants \(C,c>0\), depending only on
\((m,\beta,\pi_{\min},\alpha_{\min},\alpha_{\max},B,r,\lambda_0)\) and the sub-Gaussian
norm of \(X\), such that for every \(\delta\in(0,1)\) and
\(n\ge c\{d_\theta+\log(1/\delta)\}\),
\begin{equation}\label{eq:epsilon_rate_robust}
\epsilon_M^{\mathrm{unif}}(n,\delta)
\le
C\sqrt{\frac{d_\theta\log(Cn)+\log(1/\delta)}{n}} .
\end{equation}
Consequently, if \(a_n\) is no larger than the right-hand side of
\eqref{eq:epsilon_rate_robust} and \eqref{eq:sample_condition_robust} holds, then
Algorithm~\ref{EMrobust} satisfies
\begin{equation}\label{eq:final_rate_robust}
\|\theta^t-\theta^\star\|_2
\le
\kappa^t\|\theta^0-\theta^\star\|_2
+
\frac{C}{1-\kappa}
\sqrt{\frac{d_\theta\log(Cn)+\log(1/\delta)}{n}}
+
\frac{a_n}{1-\kappa}
\end{equation}
with probability at least \(1-\delta\).
\end{theorem}

\begin{remark}[Interpretation]
The bound \eqref{eq:final_rate_robust} separates three effects: the optimization error
\(\kappa^t\|\theta^0-\theta^\star\|_2\), the statistical error of the sample EM operator,
and the numerical error \(a_n\) from the generalized M-step. The required number of outer
iterations is the number needed to make the geometric term no larger than the
sample-plus-numerical radius:
\[
T
\gtrsim
\log_{1/\kappa}
\left(
\frac{\|\theta^0-\theta^\star\|_2}
{\sqrt{\{d_\theta\log(Cn)+\log(1/\delta)\}/n}+a_n}
\right).
\]
This is the same balancing rule used in the population-to-sample EM analysis of
\citet{balakrishnan2017statistical}: the iteration count is logarithmic in the inverse
statistical radius, and it reduces to order \(\log n\) only when \(d_\theta\), \(\delta\),
and the initialization radius are treated as fixed and \(a_n\) is no larger than the
statistical radius. Running the inner optimizer beyond this radius does not improve the
statistical order of the final estimator, although it may still be useful for numerical
diagnostics.
The proofs of Theorems~\ref{thm:robust_population_contraction},
\ref{thm:robust_sample_em}, and \ref{thm:robust_rate} are given in
\ref{app:em_population_sample_guarantees}.
\end{remark}

\section{Simulation studies} \label{sec4simulate}
This section evaluates the finite-sample behavior of the proposed log-truncated-score
mixture-regression estimator under the two-component regression design. The goal is
not only to check whether the proposed likelihood performs well under its own data
generating density, but also to assess its behavior under tail-thickened residual
regimes, compared with the pseudo-likelihoods induced by Huber, Tukey, and trimmed losses.

\subsection{Experimental design}
We use the two-component mixture regression model
\[
Y =
\begin{cases}
X^\top b_1 + \epsilon_1, & \pi_1=0.25,\\
X^\top b_2 + \epsilon_2, & \pi_2=0.75,
\end{cases}
\]
where \(b_1=(1,1)^\top\), \(b_2=(-1,-1)^\top\), \(X\sim N(0,I_2)\), and
\(\epsilon_1,\epsilon_2\) are i.i.d. In addition to a standard Gaussian baseline and the
matched case in which the errors are drawn from the proposed Catoni density itself
(sampled by the rejection sampler of~\ref{app:sampling}), the primary comparison reported below
concentrates on three non-Gaussian families: generalized normal, Laplace, and Pareto errors. These families
allow us to examine smooth symmetric tail thickening and one-sided extreme residuals without
mixing the primary display with qualitatively different stress-test mechanisms.

For the generalized normal cases, we use the type-I generalized normal, also called the
exponential-power distribution, with density
\[
f_{\mathrm{GN}}(e;\mu,s,p)
=
\frac{p}{2s\Gamma(1/p)}
\exp\left\{-\left(\frac{|e-\mu|}{s}\right)^p\right\},
\qquad s>0,\quad p>0.
\]
We set \(\mu=0\), \(s=1\), and report \(p\in\{1.25,1.50,1.75\}\).
Under this parameterization, \(p=1\) gives the Laplace distribution and \(p=2\)
gives a normal distribution with variance \(s^2/2\). Hence, the varying $p$'s give
a wide family of smooth symmetric errors with tails between Laplace and Gaussian.
The scale \(s\) is held fixed across these generalized-normal cases rather than
rescaled to equalize their variances.

We compare five estimators: the proposed Catoni-score EM estimator with fixed
\(\beta=2\), Gaussian EM, Huber pseudo-likelihood EM with tuning constant \(1.345\),
Tukey bisquare pseudo-likelihood EM with tuning constant \(4.685\), and trimmed
Gaussian EM with trimming proportion \(\tau=0.10\). The main screen reports the
theoretically motivated fixed-\(\beta\) Catoni version with \(\beta=2\). A separate
shape-sensitivity check profiles \(\beta\) over a small grid in
Subsection~\ref{subsec:beta_sensitivity}.  For each estimator and simulated dataset, we use the same
multi-start initialization pool: 20 \(k\)-means initializations on standardized
\((X,Y)\) and 8 random-subset OLS initializations. The best start is selected by the
method-specific objective value.

For every distribution, we use \(n=1000\) and \(S=100\) Monte Carlo replications. Since
labels are not identifiable in mixture models, we align estimated components to the true
components by minimizing the squared coefficient error. The reported metric is
\[
\mathrm{MSE}^{(s)}
 =
\min_{\tau\in\mathcal{S}_2}
\frac{1}{4}\sum_{j=1}^2
\left\|\widehat b_{\tau(j)}^{(s)}-b_j\right\|_2^2,
\]
where \(\mathcal{S}_2\) is the set of permutations of two component labels. The
primary table focuses on generalized-normal shapes between the Laplace and Gaussian
endpoints, high-scale Laplace settings \(s=6,\ldots,10\), and a moderate finite-variance
Pareto range.

\begin{table}[H]
\centering
\caption{Family-focused simulation results under \(n=1000\) and 100
replications. Entries are
\(1000\times\) mean coefficient MSE, with Monte Carlo standard errors in parentheses.
The smallest value in each row is bolded. The displayed grid uses generalized-normal
shapes \(p\in\{1.25,1.50,1.75\}\), Laplace scales \(s=6,\ldots,10\), and Pareto
tail indices \(a\in\{2.3,2.4,2.6,2.8,3.0\}\).}
\label{tab:sim-moderate-symmetric}
\scriptsize
\setlength{\tabcolsep}{2.2pt}
\resizebox{\textwidth}{!}{%
\begin{tabular}{lccccc}
\toprule
Error distribution & Catoni & Gaussian & Huber & Tukey & Trimmed \\
\midrule
Normal & \textbf{3.87 (0.27)} & 3.92 (0.27) & 4.50 (0.33) & 5.50 (0.42) & 50.82 (18.69) \\
Catoni-generated & \textbf{5.46 (0.41)} & 5.66 (0.43) & 6.22 (0.47) & 7.88 (0.59) & 64.56 (23.60) \\
\addlinespace
Gen. normal \(p=1.25\) & 3.99 (0.31) & 5.24 (0.55) & \textbf{3.73 (0.30)} & 4.86 (0.46) & 7.14 (0.67) \\
Gen. normal \(p=1.50\) & \textbf{2.92 (0.26)} & 3.07 (0.28) & 3.05 (0.25) & 3.11 (0.23) & 25.07 (18.17) \\
Gen. normal \(p=1.75\) & \textbf{1.98 (0.15)} & 2.04 (0.16) & 2.13 (0.18) & 2.51 (0.20) & 11.02 (3.03) \\
\addlinespace
Laplace \(s=6\) & \textbf{993.68 (40.52)} & 1014.99 (42.42) & 999.29 (43.60) & 1041.56 (57.41) & 1495.84 (149.53) \\
Laplace \(s=7\) & \textbf{1032.29 (42.92)} & 1065.55 (42.39) & 1050.19 (50.44) & 1177.37 (86.78) & 1496.20 (121.44) \\
Laplace \(s=8\) & 1065.19 (48.75) & 1105.41 (48.88) & \textbf{1046.99 (46.93)} & 1087.46 (53.90) & 2442.21 (523.06) \\
Laplace \(s=9\) & 1107.73 (54.84) & 1144.45 (57.38) & \textbf{1046.12 (53.68)} & 1357.62 (105.28) & 2180.76 (244.28) \\
Laplace \(s=10\) & 1166.18 (62.66) & 1217.13 (65.40) & \textbf{1118.64 (63.69)} & 1453.05 (128.52) & 2308.47 (272.16) \\
\addlinespace
Pareto \(a=2.3\) & \textbf{718.83 (249.99)} & 1646.45 (722.90) & 795.69 (12.61) & 750.31 (11.87) & 950.31 (50.47) \\
Pareto \(a=2.4\) & \textbf{314.43 (59.51)} & 426.96 (65.59) & 773.80 (10.10) & 767.09 (9.28) & 950.47 (46.15) \\
Pareto \(a=2.6\) & \textbf{131.60 (21.16)} & 253.41 (60.39) & 774.52 (8.54) & 783.78 (9.64) & 1006.21 (52.70) \\
Pareto \(a=2.8\) & \textbf{110.33 (27.14)} & 159.42 (42.60) & 748.55 (7.92) & 771.07 (9.30) & 859.75 (43.49) \\
Pareto \(a=3.0\) & \textbf{78.56 (15.10)} & 97.72 (26.20) & 790.44 (10.37) & 832.09 (11.99) & 1053.61 (60.84) \\
\bottomrule
\end{tabular}
}
\end{table}
Table~\ref{tab:sim-moderate-symmetric} presents the reported comparison by error family.
The proposed estimator is correctly
favored in the matched Catoni-generated case: its coefficient MSE is
\(5.46\times 10^{-3}\), compared with \(5.66\times 10^{-3}\) for Gaussian EM,
\(6.22\times 10^{-3}\) for Huber EM, and \(7.88\times 10^{-3}\) for Tukey bisquare EM.
It is also best under normal errors.

The generalized-normal grid makes the symmetric-tail transition visible. At
\(p=1.25\), Huber EM is slightly better than fixed-\(\beta\) Catoni EM. As the tails
move closer to Gaussian at \(p=1.50\) and \(p=1.75\), Catoni EM becomes the best of
the five methods, with
MSEs \(2.92\times10^{-3}\) and \(1.98\times10^{-3}\), respectively. This progression
is consistent with the proposed likelihood targeting smooth, moderately rather than
extremely heavy symmetric tails.

The reported high-scale Laplace grid is a deliberately challenging check. Catoni EM
is rowwise best at \(s=6\) and \(s=7\), but Huber EM is best for \(s=8,9,10\).
The differences among Catoni, Gaussian, and Huber EM are modest relative to their
Monte Carlo standard errors. Thus these reported settings do not establish a monotone
large-\(s\) advantage for the proposed smooth likelihood.

For the reported one-sided Pareto cases, Catoni EM has the smallest mean MSE for
\(a=2.3,2.4,2.6,2.8,\) and \(3.0\), although the result at \(a=2.3\) has
substantial Monte Carlo uncertainty. At \(a=3.0\), for example, Catoni EM reduces the
MSE from \(97.72\times 10^{-3}\) for Gaussian EM to \(78.56\times 10^{-3}\). These
asymmetric cases are not the central target setting, but they indicate useful behavior
under continuous moderately heavy one-sided tails. Appendix~\ref{app:student_t_z_estimator}
separately isolates the corresponding \(m=1\), intercept-only location-estimation
problem under symmetric Student-\(t\) errors. In that special case the
latent-allocation step disappears, and the fixed-\(\alpha,\beta\) version is a
scalar log-truncated \(Z\)-estimator rather than an EM algorithm.

This behavior is consistent with the tail shape of the proposed density. For
\(\beta=2\), the Catoni-score
\(\psi_\beta(u)=\sign(u)\log(1+|u|+|u|^\beta/\beta)\) is odd (antisymmetric about 0), and the loss \(l(u)\asymp |u|\log|u|\) as \(|u|\to\infty\).
Therefore, the induced density has heavier tails than a Gaussian density, but its tail is
similar to a Laplace density up to a logarithmic factor, and the score grows logarithmically rather than being
hard-trimmed. The proposed estimator consequently downweights large
residuals relative to Gaussian EM, but it does not fully reject gross residual outliers.
This explains why it is competitive for matched or moderately heavy symmetric errors.
Accordingly, we view the Catoni-score likelihood as a smooth efficiency-robustness
compromise rather than a universal replacement for estimators with bounded or
redescending influence.

\subsection{Sensitivity to the shape parameter}\label{subsec:beta_sensitivity}

The main simulation fixes the Catoni shape parameter at
\(\beta=2\). This is the pre-specified version tied to the density construction and to
the default Catoni-type score used in the algorithmic and theoretical sections. It also
avoids treating a weakly identified shape parameter as an extra tuning parameter in the
main comparison. Indeed, after the component scales have been profiled out, the
likelihood can be relatively flat in \(\beta\), and profile estimates may concentrate
at the admissible boundary. For this reason, the fixed-\(\beta=2\) estimator is our
main estimator, while the profile-likelihood estimation of \(\beta\) can be treated as a
sensitivity analysis.

To examine whether a non-leading fixed-\(\beta\) result is caused by the fixed shape,
we apply the profiled-shape version to the generalized-normal \(p=1.25\) row and to
the high-scale Laplace rows
\(s=8,9,10\), for which fixed-\(\beta\) EM does not lead in the primary
table. In each Monte Carlo replication, we use the same generated sample, seed, and
initialization pool as in Table~\ref{tab:sim-moderate-symmetric}, and evaluate the
common-\(\beta\) profile likelihood on the pre-specified grid
\[
\mathcal B=\{1.10,1.25,1.50,1.75,2.00\}.
\]
For a fixed candidate \(\beta\), define the profiled Catoni observed log-likelihood by
\begin{equation}\label{eq:catoni_profile_likelihood_sim}
\ell_{\rm Catoni}(\beta)
=
\max_{(\pi,b,\alpha):\,(\pi,b,\alpha,\beta)\in\Omega_\beta}
\sum_{j=1}^n
\log\left\{
\sum_{i=1}^m
\pi_i f(Y_j-X_j^\top b_i;\alpha_i,\beta)
\right\}.
\end{equation}
The maximization in \eqref{eq:catoni_profile_likelihood_sim} is carried out with the same
constrained GEM routine and initialization pool used for the fixed-\(\beta\) fits. The
profile estimator uses
\[
\widehat{\beta}_{\rm PL}
=
\arg\max_{\beta\in\mathcal B}
\ell_{\rm Catoni}(\beta).
\]
It is therefore selected by observed likelihood rather
than by coefficient MSE. The lower endpoint \(1.10\) is a conservative
pre-specified bound rather than a claim that the optimum must occur away from one.

\begin{table}[htbp]
\centering
\caption{Profile-likelihood sensitivity of the Catoni estimator to the common shape
parameter \(\beta\).
Entries are \(1000\times\) mean coefficient MSE with Monte Carlo standard errors in
parentheses, using the same \(n=1000\) and 100 replications as
Table~\ref{tab:sim-moderate-symmetric}. The Catoni
\((\beta=\widehat\beta_{\rm PL})\) column maximizes the Catoni observed likelihood over
\(\beta\in\{1.10,1.25,1.50,1.75,2.00\}\) within each replication. The
\(\beta\)-selection column reports the modal profile estimate and its frequency; a modal
value at \(1.10\) indicates lower-end selection on this pre-specified grid. The
\(\widehat\pi\) column reports Monte Carlo mean fitted mixing weights for
Catoni \((\beta=\widehat\beta_{\rm PL})\) after label alignment; the true weights are
\((0.25,0.75)\). Boldface
marks the smallest reported MSE in each row.}
\label{tab:beta-sensitivity}
\scriptsize
\setlength{\tabcolsep}{3pt}
\resizebox{\textwidth}{!}{%
\begin{tabular}{lccccc}
\toprule
Error distribution & Catoni \((\beta=2)\) & Catoni \((\beta=\widehat\beta_{\rm PL})\) & Selected \(\beta\) (mode) & Mean \(\widehat\pi\) & Best non-Catoni method \\
\midrule
Gen. normal \(p=1.25\) & 3.99 (0.31) & \textbf{3.41 (0.27)} & 1.10 (100/100) & (0.259, 0.741) & Huber: 3.73 (0.30) \\
Laplace \(s=8\) & 1065.19 (48.75) & \textbf{1035.55 (50.27)} & 1.10 (91/100) & (0.497, 0.503) & Huber: 1046.99 (46.93) \\
Laplace \(s=9\) & 1107.73 (54.84) & 1087.91 (57.35) & 1.10 (90/100) & (0.495, 0.505) & \textbf{Huber: 1046.12 (53.68)} \\
Laplace \(s=10\) & 1166.18 (62.66) & 1140.92 (64.94) & 1.10 (90/100) & (0.498, 0.502) & \textbf{Huber: 1118.64 (63.69)} \\
\bottomrule
\end{tabular}
}
\end{table}
Table~\ref{tab:beta-sensitivity} shows that the fixed value \(\beta=2\) can be
conservative. In the heavier generalized-normal setting, estimating \(\beta\) by
likelihood reduces the MSE from \(3.99\times10^{-3}\) to \(3.41\times10^{-3}\),
overtaking Huber EM. In the high-scale Laplace settings, it reduces the fixed-Catoni
MSE from \(1065.19\), \(1107.73\), and \(1166.18\) to \(1035.55\), \(1087.91\),
and \(1140.92\), respectively, on the \(1000\times\)MSE scale. For \(s=8\), the estimated mean by $\widehat\beta_{\rm PL}$ is slightly below Huber EM. For \(s=9\) and \(s=10\),
Huber EM remains lower in mean MSE.

The modal profile estimate is \(1.10\) in all four diagnostic rows, occurring in
100/100 generalized-normal replications and in 91/100, 90/100, and 90/100 Laplace
replications. The fitted mixing weights remain close to the true \((0.25,0.75)\) in the
generalized-normal row, but the high-scale Laplace rows move toward balanced component
weights, illustrating that the estimated shape can improve coefficient MSE without
recovering all mixture parameters equally well. Mechanistically, smaller \(\beta\) makes
the large-residual score grow
more slowly, so the fitted likelihood gives less leverage to tail observations than
the fixed-\(\beta=2\) version. However, lower-end profile selection is not the same as
proving that ``smaller is always better.'' The shape is estimated by likelihood,
whereas the reported performance metric is coefficient MSE. These criteria need not
have the same optimum: reducing \(\beta\) increases robustness to large residuals, but
it can also reduce the effective information in the estimating equations. Additional
pilot checks with a denser lower grid showed that the likelihood can move below
\(1.10\) while the coefficient MSE changes only marginally and can even worsen.

Thus the sensitivity analysis supports the main interpretation of the method:
estimating the common shape can benefit the Catoni-score
likelihood when the fixed shape imposes insufficient downweighting. It improves all
four examined non-leading fixed-\(\beta\) rows, but does not establish uniform
dominance over Huber EM under high-scale Laplace errors. The  \(\beta=2\)
version remains the pre-specified estimator of record in the primary comparison, and
 \(\beta=\widehat\beta_{\rm PL}\) should be read as a sensitivity extension.

\section{Real data applications}\label{sec:real}
\subsection{French motor insurance claim severity}\label{sec5real}

The freMTPL French motor third-party liability portfolio
\citep{dutang2025casdatasets, charpentier2014computational} provides a strongly
right-skewed severity benchmark. We model the 16,181 recorded positive individual claim
payments, matched to policy covariates. It deliberately do not model the zero-claim
frequency process. The response is the individual payment amount, and the regression
uses an intercept with \texttt{Exposure}, \texttt{CarAge}, \texttt{DriverAge}, and
\texttt{Density}; implementation details and split construction are given in ~\ref{app:realdata_details}. The mean payment is 2,129.97
euros, the median is 1,156.00 euros, and the largest observed payment is 2,036,833.00
euros, which show  that this dataset has many outliers.

\begin{figure}[htbp]
\centering
\includegraphics[width=0.95\textwidth]{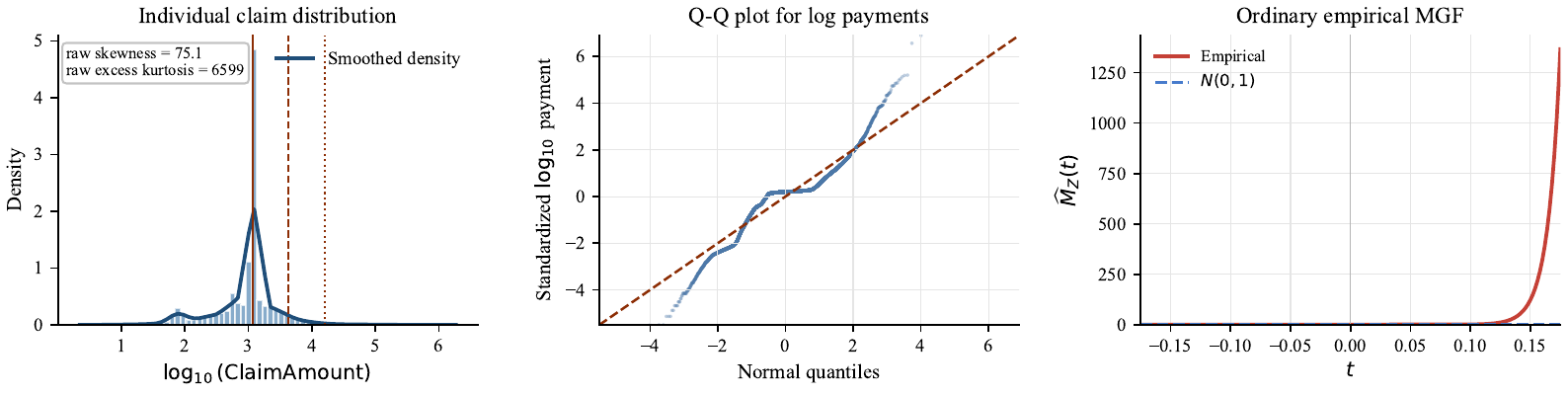}
\caption{Distribution and tail trend of individual positive freMTPL claim payments.
The left panel shows a histogram of \(\log_{10}(\texttt{ClaimAmount})\) with a smoothed
density curve; the vertical lines mark the median, 95th percentile, and 99th percentile.
The middle panel gives a normal Q-Q plot for standardized log payments. The right
panel shows the empirical MGF of standardized raw payments,
\(\widehat M_Z(t)=n^{-1}\sum_i\exp(tZ_i)\), against the standard-normal benchmark.}
\label{fig:fremtpl-claim-distribution}
\end{figure}

Figure~\ref{fig:fremtpl-claim-distribution} visualizes why this portfolio is a useful
real-data stress test for a residual-robust likelihood. The raw payments have skewness
75.1 and excess kurtosis above 6500, and the log-payment Q-Q plot still shows a strong
upper-tail departure from Gaussianity.

We therefore model the \emph{log}-severity \(\log(\texttt{ClaimAmount})\), the standard
actuarial scale, on which the response is near-symmetric and only moderately heavy (skewness
\(-0.59\), excess kurtosis \(3.42\)). It matches the regime the proposed likelihood targets.
All five estimators are fit with the same generalized-EM updates used in
Section~\ref{sec4simulate}, on a \(60/20/20\) policy-grouped split stratified by
maximum-claim decile (\(3{,}247\) test claims); the shape parameter is selected by training-set
profile likelihood, giving \(\widehat\beta=1.75\).
Table~\ref{tab:fremtpl-logseverity-coefficients} reports the fitted coefficients and
Table~\ref{tab:fremtpl-logseverity-predictive} the held-out errors.

\begin{table}[htbp]
\centering
\caption{Fitted two-component coefficients on the freMTPL \emph{log}-severity training set.
Numeric covariates are standardized on the training split; \(\widehat\pi_j\) are mixing
weights and the scale column is \(\alpha\) for Catoni and \(\sigma\) for the other methods.
On the log scale the Catoni mixture is balanced and interpretable, unlike the degenerate
raw-euro fits.}
\label{tab:fremtpl-logseverity-coefficients}
\scriptsize
\setlength{\tabcolsep}{4pt}
\resizebox{\textwidth}{!}{%
\begin{tabular}{lcccccccc}
\toprule
Method & Comp.\ & \(\widehat\pi_j\) & Scale & Intercept & Exposure & CarAge & DriverAge & Density \\
\midrule
Catoni EM \((\widehat\beta=1.75)\) & 1 & 0.545 & 1.326 & 6.634 & -0.250 & -0.097 & 0.066 & -0.033 \\
 & 2 & 0.455 & 0.042 & 7.066 & -0.002 & -0.001 & 0.002 & 0.003 \\
Gaussian EM & 1 & 0.546 & 1.457 & 6.618 & -0.253 & -0.101 & 0.066 & -0.035 \\
 & 2 & 0.454 & 0.044 & 7.066 & -0.002 & -0.001 & 0.002 & 0.003 \\
Huber EM & 1 & 0.528 & 1.382 & 6.621 & -0.265 & -0.105 & 0.072 & -0.034 \\
 & 2 & 0.472 & 0.048 & 7.066 & -0.003 & -0.001 & 0.002 & 0.003 \\
Tukey EM & 1 & 0.100 & 0.942 & 6.601 & -0.375 & -0.065 & 0.100 & -0.043 \\
 & 2 & 0.900 & 0.129 & 7.070 & -0.006 & -0.002 & 0.002 & 0.002 \\
Trimmed Gaussian EM & 1 & 0.510 & 0.925 & 6.898 & -0.164 & -0.082 & 0.021 & -0.043 \\
 & 2 & 0.490 & 0.043 & 7.066 & -0.002 & -0.001 & 0.002 & 0.003 \\
\bottomrule
\end{tabular}
}
\end{table}

\begin{table}[htbp]
\centering
\caption{Held-out predictive comparison for freMTPL \emph{log}-severity. RMSE, MAE, and
MedAE are on the \(\log\)-euro modeling scale; MSE is in squared \(\log\)-euro. Lower is
better; the column minimum is bold. The Catoni likelihood gives the best RMSE and beats the
Gaussian likelihood on every metric, while high-breakdown trimming attains a lower central
error at a higher RMSE.}
\label{tab:fremtpl-logseverity-predictive}
\small
\setlength{\tabcolsep}{6pt}
\begin{tabular}{llcccc}
\toprule
Method & Shape or tuning & RMSE & MSE & MAE & MedAE \\
\midrule
Catoni EM \((\widehat\beta=1.75)\) & -- & \textbf{1.1264} & \textbf{1.2688} & 0.7465 & 0.3892 \\
Gaussian EM & -- & 1.1266 & 1.2692 & 0.7501 & 0.3951 \\
Huber EM & \(c=1.345\) & 1.1266 & 1.2692 & 0.7468 & 0.3915 \\
Tukey EM & \(c=4.685\) & 1.1441 & 1.3090 & \textbf{0.6720} & \textbf{0.1749} \\
Trimmed Gaussian EM & \(\tau=0.10\) & 1.1370 & 1.2928 & 0.6928 & 0.2356 \\
\bottomrule
\end{tabular}
\end{table}

On the log scale the Catoni \((\widehat\beta=1.75)\) fit recovers a balanced, interpretable two-component
mixture (\(\widehat\pi=(0.55,0.45)\), log-euro scales \(1.33\) and \(0.04\)), in contrast to
the degenerate raw-euro decomposition. It attains the lowest held-out RMSE and improves on
the Gaussian likelihood in every column, while being essentially tied with Huber EM on mean
and median absolute error. Consistent with the method's scope, the high-breakdown Tukey and
trimmed procedures attain a smaller central error by discarding upper-tail observations, at
the cost of a higher RMSE. The proposed likelihood is thus a coherent and competitive
smooth-likelihood model for log-severity, rather than a high-breakdown replacement for
trimming under the raw one-sided response.

\subsection{Crypto return innovations}\label{sec:crypto_realdata}

The insurance example above is deliberately difficult for a symmetric additive
residual model because individual claim payments are positive and extremely
right-skewed. We therefore add a second real-data experiment whose data-generating
features are closer to the robustness scope in Section~\ref{subsec:influence_scope}:
centered, two-sided return innovations with occasional residual spikes. We use public
Binance Spot kline data for \texttt{BTCUSDT}, \texttt{ETHUSDT}, \texttt{BNBUSDT}, and
\texttt{SOLUSDT} from 2021--2024. Here BTC, ETH, BNB, and SOL denote Bitcoin, Ether,
the BNB token (formerly Binance Coin), and Solana, respectively, and USDT denotes the
Tether quote currency. Let
\(P_t^{(a)}\) denote the close price of asset
\(a\). The response is the ETH close-to-close log return,
\[
r_t^{(a)}=\log P_t^{(a)}-\log P_{t-1}^{(a)}
\]
in basis points, and the two-component regression uses contemporaneous BTC, BNB, and
SOL returns together with lagged BTC:
\[
\begin{aligned}
Y_t&=r_t^{(\mathrm{ETH})},\\
X_t&=\left(1,\ r_t^{(\mathrm{BTC})},\ r_{t-1}^{(\mathrm{BTC})},\
r_t^{(\mathrm{BNB})},\ r_t^{(\mathrm{SOL})}\right),\\
Y_t&=
b_{0,z_t}+b_{1,z_t}r_t^{(\mathrm{BTC})}
+b_{2,z_t}r_{t-1}^{(\mathrm{BTC})}
+b_{3,z_t}r_t^{(\mathrm{BNB})}
+b_{4,z_t}r_t^{(\mathrm{SOL})}
+\epsilon_{t,z_t},\qquad z_t\in\{1,2\}.
\end{aligned}
\]
The four non-intercept covariates are standardized using the training split only.
The clean 5-minute experiment comprises 420,539 regression observations, split
chronologically into 252,323 training, 84,108 validation, and 84,108 test observations;
additional data-processing details are given in \ref{app:binance_details}. The 5-minute training response has standard
deviation 30.77 basis points and excess kurtosis 140.85, with a maximum absolute return
of 1646.27 basis points.

\begin{figure}[htbp]
\centering
\includegraphics[width=0.95\textwidth]{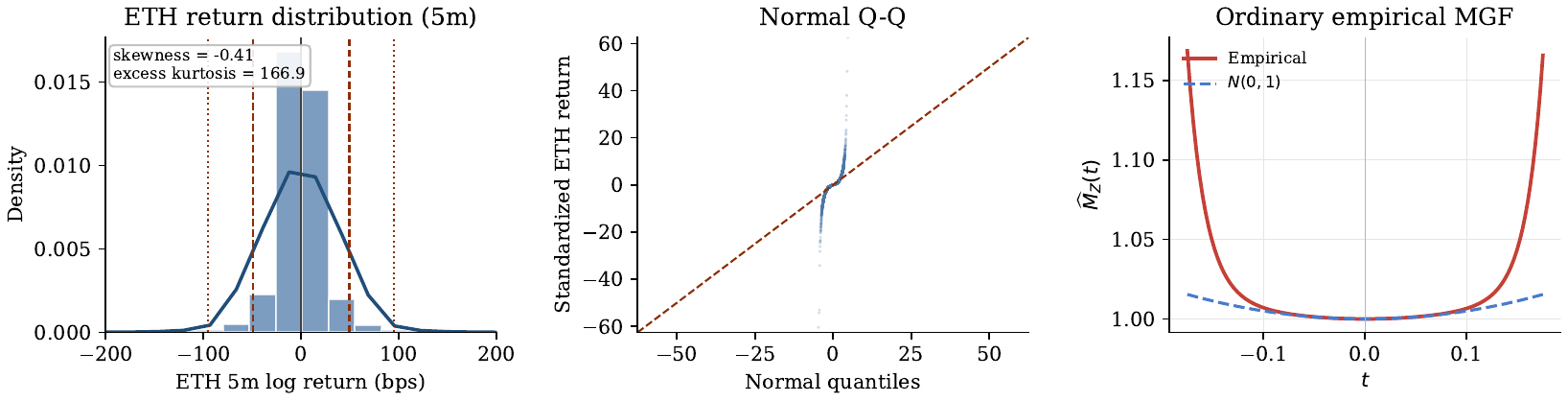}
\caption{Distribution and tail trend of five-minute ETH return innovations in the
Binance crypto experiment. The left panel gives the ETH return histogram in basis points
with a smoothed density curve, the median, and symmetric 95th and 99th absolute-return
cutoffs; the histogram is zoomed to \([-200,200]\) basis points, while the full-sample
maximum absolute return is reported in the text. The middle panel gives a normal Q-Q
plot for standardized ETH returns. The right panel shows the ordinary empirical MGF of
standardized ETH returns,
\(\widehat M_Z(t)=n^{-1}\sum_i\exp(tZ_i)\), against the standard-normal benchmark.}
\label{fig:binance-crypto-diagnostics}
\end{figure}

Table~\ref{tab:binance-predictive-results} reports the 5-minute held-out results from
the full training split. The Catoni \((\widehat\beta=1.05)\) fit selects the lower endpoint
of the profile-likelihood grid, as in the 1-minute fit, indicating that the shape
selection favors the most logarithmic score in the candidate grid. This Catoni
\((\widehat\beta=1.05)\) fit gives the
lowest MAE and median absolute error. The
differences are small in absolute value: relative to Gaussian EM, the 5-minute Catoni
\((\widehat\beta=1.05)\) fit reduces MAE from 6.5152 to 6.5104 basis points.
Huber EM has the lowest RMSE and MSE, but its MAE and median absolute error are larger.
\begin{table}[htbp]
\centering
\caption{Held-out predictive comparison for 5-minute Binance crypto return innovations.
RMSE, MAE, and MedAE are reported in basis points; MSE in squared basis points.
Lower values are better.}
\label{tab:binance-predictive-results}
\small
\setlength{\tabcolsep}{5pt}
\begin{tabular}{llcccc}
\toprule
Method & Shape or tuning & RMSE & MSE & MAE & MedAE \\
\midrule
Catoni EM \((\beta=2.00)\) & -- & 10.0091 & 100.1825 & 6.5132 & 4.5032 \\
Catoni EM \((\widehat\beta=1.05)\) & -- & 10.0060 & 100.1195 & \(\mathbf{6.5104}\) & \(\mathbf{4.5023}\) \\
Gaussian EM & -- & 10.0062 & 100.1241 & 6.5152 & 4.5093 \\
Huber EM & \(c=1.345\) & \(\mathbf{9.9996}\) & \(\mathbf{99.9926}\) & 6.5409 & 4.5495 \\
Tukey EM & \(c=4.685\) & 10.0111 & 100.2218 & 6.5248 & 4.5230 \\
Trimmed Gaussian EM & \(\tau=0.10\) & 10.0067 & 100.1336 & 6.5247 & 4.5248 \\
\bottomrule
\end{tabular}
\end{table}
Table~\ref{tab:binance-catoni-coefficients} gives the fitted 5-minute coefficients for
all methods from the full training split; we focus on the Catoni \((\widehat\beta=1.05)\)
fit. Component labels are ordered by the fitted
component mean on the training set. The larger component has weight 0.7499 and scale
6.69 basis points, while the smaller component has weight 0.2501 and scale 21.99 basis
points. The smaller component therefore behaves like a higher-volatility regime. Its
standardized contemporaneous BTC, lagged BTC, and SOL coefficients are larger than in
the low-scale component, while both components preserve positive contemporaneous
cross-asset slopes.

\begin{table}[htbp]
\centering
\caption{Fitted 5-minute two-component coefficients for the Binance experiment, in basis
points on the train-standardized covariate scale. \(\widehat\pi_j\) are mixing weights and
the scale column is \(\alpha\) for Catoni and \(\sigma\) for the other methods; component
labels are ordered by the fitted component mean.}
\label{tab:binance-catoni-coefficients}
\scriptsize
\setlength{\tabcolsep}{4pt}
\resizebox{\textwidth}{!}{%
\begin{tabular}{lcccccccc}
\toprule
Method & Comp.\ & \(\widehat\pi_j\) & Scale & Intercept & BTC\(_t\) & BTC\(_{t-1}\) & BNB\(_t\) & SOL\(_t\) \\
\midrule
Catoni EM \((\beta=2.00)\) & 1 & 0.7683 & 6.3675 & -0.1122 & 18.3646 & -0.3309 & 6.0795 & 3.3036 \\
 & 2 & 0.2317 & 22.8331 & 0.4021 & 20.0503 & 0.6308 & 7.0312 & 4.6146 \\
Catoni EM \((\widehat\beta=1.05)\) & 1 & 0.7499 & 6.6928 & -0.1262 & 17.9970 & -0.5182 & 6.1495 & 2.9664 \\
 & 2 & 0.2501 & 21.9888 & 0.3983 & 20.8397 & 1.1134 & 6.7849 & 5.6302 \\
Gaussian EM & 1 & 0.7989 & 7.1575 & -0.1105 & 18.4385 & -0.2651 & 6.2186 & 3.3632 \\
 & 2 & 0.2011 & 27.8798 & 0.5477 & 19.9670 & 0.3808 & 7.2459 & 4.4125 \\
Huber EM & 1 & 0.5017 & 6.2534 & -0.2562 & 15.0492 & -2.0789 & 2.6608 & 5.3647 \\
 & 2 & 0.4983 & 8.5506 & 0.2560 & 22.8181 & 2.2546 & 11.5278 & 1.4283 \\
Tukey EM & 1 & 0.5401 & 7.0578 & -0.1531 & 15.3199 & -0.7950 & 3.4099 & 5.3534 \\
 & 2 & 0.4599 & 7.8561 & 0.0301 & 23.0107 & 0.7265 & 10.3797 & 1.1336 \\
Trimmed Gaussian EM & 1 & 0.5062 & 6.1210 & -0.2019 & 15.8089 & -1.9174 & 2.8362 & 5.1469 \\
 & 2 & 0.4938 & 6.7607 & 0.0468 & 21.7536 & 1.8927 & 10.9885 & 1.3229 \\
\bottomrule
\end{tabular}
}
\end{table}

The training-only contamination sensitivity is reported in
~\ref{app:binance_details}. Its message is consistent with the clean comparison:
the Catoni likelihood is a coherent and competitive residual-innovation model relative
to Gaussian EM, but it does not dominate Huber, Tukey, or trimmed procedures under all
tail-sensitive metrics.

\section{Conclusions}\label{sec6}

We introduced a log-truncated likelihood and a constrained generalized EM
algorithm for mixture regression. The defensible novelty is the coherent use of a
logarithmically growing residual score in likelihood-based responsibilities, regression
updates, and scale updates. The induced error density is light-tailed: its log density
decays at order \(-|x|\log |x|\), and the robustness claim concerns residual-score
growth rather than tail thickness of the fitted law.

The empirical study is designed to evaluate the method as a residual-robust likelihood,
not as a high-breakdown leverage-robust procedure. In simulations, the method is
competitive for matched Catoni-generated errors, generalized-normal errors between
Laplace and Gaussian shapes, high-scale Laplace errors, and finite-variance Pareto
errors. The profile-shape extension can improve performance in some tail-thickened
settings, although boundary selections indicate weak shape identification. In real
data, the French motor insurance severity example
\citep{dutang2025casdatasets, charpentier2014computational} yields a balanced,
interpretable two-component mixture on the standard log-severity scale --- a
near-symmetric, moderately heavy regime the method targets --- and the Binance return
experiment shows competitive likelihood and typical-error performance for centered
two-sided residual innovations. Huber, Tukey, or trimmed procedures may still be
preferable under metrics emphasizing hard rejection or tail-specific error.

The theory and experiments consequently support a deliberately limited robustness
claim. Relative to Gaussian mixture regression, the proposed estimator reduces the
growth of the residual score, but its influence remains unbounded and it does not
protect against arbitrary high-leverage covariates. The conditional local GEM bound
also requires bounded-design or corresponding design-tail conditions. The current
empirical results illustrate useful residual-outlier behavior, but broader comparisons
with Student-\(t\), Laplace, contaminated-normal, and high-breakdown mixture regression
methods, together with a full frequency-severity insurance study, are needed before
making stronger comparative claims.

\clearpage
\appendix
\renewcommand{\thetheorem}{\Alph{section}.\arabic{theorem}}
\renewcommand{\thelemma}{\Alph{section}.\arabic{lemma}}
\renewcommand{\theproposition}{\Alph{section}.\arabic{proposition}}

\section{One-component Catoni-type \texorpdfstring{\(Z\)}{Z}-estimation}
\label{app:student_t_z_estimator}

This appendix records the scalar location problem that motivates the robust
score used in the mixture-regression likelihood. It should be read as the
one-component specialization of Algorithm~\ref{EMrobust}: there are no latent
responsibilities, no mixing-weight update, and the regression block reduces to
a scalar \(Z\)-estimating equation.
Let
\[
  \xi_i=\mu+\sigma\varepsilon_i,\qquad i=1,\ldots,n.
\]
The errors need not be assumed to follow a Student-\(t\) law. For the population
location interpretation below, it is enough that they are centered and
symmetric, and that \(\mathbb E|\varepsilon_i|^{1+\eta}<\infty\) for some
\(\eta>0\); this finite-moment condition implies the integrability of the
logarithmic score. Student-\(t_\nu\) errors with \(\nu>1\) are only one
representative example used in the numerical check.

For \(1<\beta\le2\), define
\[
\psi_\beta(x)=\operatorname{sign}(x)
\log\left(1+|x|+\frac{|x|^\beta}{\beta}\right),
\qquad
A_\beta(z)=\int_0^z
\log\left(1+u+\frac{u^\beta}{\beta}\right)\,du .
\]
The scalar working log-likelihood corresponding to \eqref{eq:proposed-density}
is
\begin{equation}\label{eq:scalar_z_loglik}
\ell_Z(\theta,\alpha,\beta)
=
\sum_{i=1}^n
\left[
-\log\alpha-\log I_\beta
-A_\beta\left(\frac{|\xi_i-\theta|}{\alpha}\right)
\right],
\qquad
\alpha>0,\quad \beta\in[\beta_{\min},2].
\end{equation}
For fixed \((\alpha,\beta)\), maximizing \eqref{eq:scalar_z_loglik} over
\(\theta\) is equivalent to solving the Catoni-type \(Z\)-equation
\begin{equation}\label{eq:scalar_z_equation}
\sum_{i=1}^n
\psi_\beta\left(\frac{\xi_i-\theta}{\alpha}\right)=0 .
\end{equation}
Because \(\psi_\beta\) is odd and strictly increasing, the left side of
\eqref{eq:scalar_z_equation} is continuous and strictly decreasing in
\(\theta\), so the root is unique and can be found by bisection.

The scale and shape blocks are the \(m=1\) analogues of the
\(\alpha/\beta\)-step in Algorithm~\ref{EMrobust}. Given an updated location
\(\theta^{(t+1)}\), write \(r_i^{(t+1)}=\xi_i-\theta^{(t+1)}\). For a fixed
\(\beta\), the scale update solves
\begin{equation}\label{eq:scalar_alpha_root}
G_Z^{(t)}(\alpha)
:=
\sum_{i=1}^n
\frac{|r_i^{(t+1)}|}{\alpha}
\log\left(
1+\frac{|r_i^{(t+1)}|}{\alpha}
+\frac{1}{\beta}
\left(\frac{|r_i^{(t+1)}|}{\alpha}\right)^\beta
\right)
-n
=0,
\end{equation}
with projection to \([\alpha_{\min},\alpha_{\max}]\). When \(\beta\) is
estimated, let \(\widehat\alpha(\beta)\) be the corresponding constrained
scale maximizer from \eqref{eq:scalar_alpha_root}. The profiled shape objective
is
\begin{equation}\label{eq:scalar_beta_profile}
Q_{Z,p}^{(t)}(\beta)
=
\sum_{i=1}^n
\left[
-\log\widehat\alpha(\beta)
-\log I_\beta
-A_\beta\left(\frac{|r_i^{(t+1)}|}{\widehat\alpha(\beta)}\right)
\right],
\qquad
\beta\in[\beta_{\min},2],
\end{equation}
and the candidate update is
\begin{equation}\label{eq:scalar_beta_update}
\beta_{\mathrm{cand}}^{(t+1)}
\in
\argmax_{\beta\in[\beta_{\min},2]} Q_{Z,p}^{(t)}(\beta).
\end{equation}
As in Algorithm~\ref{EMrobust}, this one-dimensional profile search can be
implemented by a grid search with optional Brent or damped Newton refinement,
and the pair
\(\{\widehat\alpha(\beta_{\mathrm{cand}}^{(t+1)}),
\beta_{\mathrm{cand}}^{(t+1)}\}\) is accepted only if
\(\ell_Z\) does not decrease; otherwise the update keeps the previous
\(\beta\) and performs the fixed-\(\beta\) scale update.

\begin{algorithm}[H]
\caption{One-component Catoni-type \(Z\)-estimation with profile \(\beta\).}
\label{alg:student_t_z_estimator}
\small
\begin{algorithmic}[1]
\REQUIRE Data \(\xi_1,\ldots,\xi_n\), fixed \(\beta\) or initial
\(\beta^{(0)}\in[\beta_{\min},2]\), constraints
\(\alpha_{\min},\alpha_{\max}\), tolerance \(\delta\), and initial
\((\theta^{(0)},\alpha^{(0)})\).
\ENSURE Location estimate \(\widehat\mu_Z\), scale estimate \(\widehat\alpha\), and shape estimate \(\widehat\beta\) when estimated.
\STATE Compute \(\ell_Z(\theta^{(0)},\alpha^{(0)},\beta^{(0)})\); set \(t=0\). If \(\beta\) is fixed, set \(\beta^{(t)}\equiv\beta\).
\REPEAT
\STATE \textbf{\(Z\)-step:} With \((\alpha^{(t)},\beta^{(t)})\) fixed, solve \eqref{eq:scalar_z_equation} by bisection and set the root to \(\theta^{(t+1)}\).
\STATE \textbf{\(\alpha/\beta\)-step:} If \(\beta\) is fixed, solve \eqref{eq:scalar_alpha_root} and clip to \([\alpha_{\min},\alpha_{\max}]\). If \(\beta\) is estimated, compute the profile candidate \eqref{eq:scalar_beta_update}, update the corresponding \(\alpha\), and accept the pair only when \(\ell_Z\) does not decrease; otherwise keep \(\beta^{(t)}\) and perform the fixed-\(\beta\) scale update.
\STATE Compute \(\ell_Z(\theta^{(t+1)},\alpha^{(t+1)},\beta^{(t+1)})\).
\STATE Set \(t\leftarrow t+1\).
\UNTIL{\(|\ell_Z(\theta^{(t)},\alpha^{(t)},\beta^{(t)})-\ell_Z(\theta^{(t-1)},\alpha^{(t-1)},\beta^{(t-1)})|\le\delta\)}
\STATE \textbf{Output:} \(\widehat\mu_Z=\theta^{(t)}\), \(\widehat\alpha=\alpha^{(t)}\), and \(\widehat\beta=\beta^{(t)}\) if estimated.
\end{algorithmic}
\end{algorithm}

At the population level, symmetry makes the location target transparent. If
\(\xi-\mu\) is symmetrically distributed, then for every fixed
\(\alpha>0\) and \(1<\beta\le2\),
\[
  \mathbb E\,
  \psi_\beta\left(\frac{\xi-\mu}{\alpha}\right)=0
\]
whenever the expectation is defined. The logarithmic growth of \(\psi_\beta\)
guarantees this integrability under the finite-\((1+\eta)\)-moment condition.
Thus the scalar root targets the same mean/location parameter \(\mu\), while
downweighting extreme observations relative to the sample mean.

Under the usual smooth \(M\)-estimation regularity conditions, the corresponding
asymptotic variance has the standard sandwich form
\[
  V(\alpha,\beta)
  =
  \alpha^2
  \frac{\mathbb E\left[\psi_\beta\{(\xi-\mu)/\alpha\}^2\right]}
  {\left(\mathbb E\left[\psi_\beta'\{(\xi-\mu)/\alpha\}\right]\right)^2},
  \qquad
  \psi_\beta'(z)=
  \frac{1+|z|^{\beta-1}}
       {1+|z|+|z|^\beta/\beta}.
\]
This variance is finite even when \(\mathbb E\varepsilon_i^2=\infty\), because
\(\psi_\beta\) grows only logarithmically. The mixture regression estimator in
the main text can therefore be viewed as the EM-compatible extension of this
one-component robust-score construction.

Table~\ref{tab:student-t-z-estimator} gives a finite-sample check for this
scalar location problem using Student-\(t\) errors as a representative
finite-mean, possibly infinite-variance example. We report the standard Monte
Carlo summaries bias, empirical standard deviation, root mean squared error,
and mean absolute error. For \(\nu\le2\), the population variance of the sample
mean is infinite, so the sample-mean SD and RMSE columns should be read only as
finite-run diagnostics. The median is a strong benchmark in this symmetric
one-dimensional setting. The \(Z\)-estimators remain nearly unbiased and stable
across the grid, and smaller \(\beta\) gives slightly stronger tail
downweighting in the heaviest cases.

\begin{table}[htbp]
\centering
\caption{Student-\(t_\nu\) location simulation for the scalar Catoni-type \(Z\)-estimator. The target is \(\mu=0\); \(n=1000\) observations are drawn from \(t_\nu\), and each entry uses 2,000 replications. The \(Z\)-estimators solve \(\sum_i\psi_\beta\{(\xi_i-\theta)/\widehat\alpha\}=0\) with \(\widehat\alpha=\operatorname{MAD}/0.6745\). Bias, SD, RMSE, and MAE are Monte Carlo summaries. For \(\nu\le2\), the sample mean has infinite population variance, so its displayed SD and RMSE are finite-simulation diagnostics rather than stable population quantities.}
\label{tab:student-t-z-estimator}
\scriptsize
\setlength{\tabcolsep}{3.4pt}
\begin{tabular}{cc l rrrr}
\toprule
\(\nu\) & \(n\) & Estimator & Bias & SD & RMSE & MAE \\
\midrule
1.2 & 1000 & Sample mean & -0.004 & 2.156 & 2.156 & 0.795 \\
 &  & Sample median & 0.002 & 0.048 & 0.048 & 0.039 \\
 &  & \(Z\)-estimator, \(\beta=1.05\) & 0.001 & 0.066 & 0.066 & 0.052 \\
 &  & \(Z\)-estimator, \(\beta=1.10\) & 0.001 & 0.067 & 0.067 & 0.053 \\
 &  & \(Z\)-estimator, \(\beta=1.25\) & 0.001 & 0.070 & 0.070 & 0.055 \\
 &  & \(Z\)-estimator, \(\beta=1.50\) & 0.001 & 0.075 & 0.075 & 0.059 \\
 &  & \(Z\)-estimator, \(\beta=1.75\) & 0.001 & 0.081 & 0.081 & 0.063 \\
 &  & \(Z\)-estimator, \(\beta=2.00\) & 0.001 & 0.087 & 0.087 & 0.068 \\
\addlinespace
1.5 & 1000 & Sample mean & 0.009 & 0.459 & 0.459 & 0.236 \\
 &  & Sample median & 0.001 & 0.048 & 0.048 & 0.039 \\
 &  & \(Z\)-estimator, \(\beta=1.05\) & 0.002 & 0.057 & 0.057 & 0.046 \\
 &  & \(Z\)-estimator, \(\beta=1.10\) & 0.002 & 0.058 & 0.058 & 0.046 \\
 &  & \(Z\)-estimator, \(\beta=1.25\) & 0.002 & 0.060 & 0.060 & 0.048 \\
 &  & \(Z\)-estimator, \(\beta=1.50\) & 0.002 & 0.063 & 0.063 & 0.050 \\
 &  & \(Z\)-estimator, \(\beta=1.75\) & 0.002 & 0.066 & 0.066 & 0.053 \\
 &  & \(Z\)-estimator, \(\beta=2.00\) & 0.002 & 0.070 & 0.070 & 0.056 \\
\addlinespace
2.0 & 1000 & Sample mean & 0.004 & 0.113 & 0.113 & 0.082 \\
 &  & Sample median & 0.002 & 0.044 & 0.044 & 0.035 \\
 &  & \(Z\)-estimator, \(\beta=1.05\) & 0.002 & 0.047 & 0.047 & 0.037 \\
 &  & \(Z\)-estimator, \(\beta=1.10\) & 0.002 & 0.047 & 0.047 & 0.038 \\
 &  & \(Z\)-estimator, \(\beta=1.25\) & 0.002 & 0.048 & 0.048 & 0.039 \\
 &  & \(Z\)-estimator, \(\beta=1.50\) & 0.003 & 0.050 & 0.050 & 0.040 \\
 &  & \(Z\)-estimator, \(\beta=1.75\) & 0.003 & 0.052 & 0.052 & 0.041 \\
 &  & \(Z\)-estimator, \(\beta=2.00\) & 0.003 & 0.054 & 0.054 & 0.043 \\
\addlinespace
3.0 & 1000 & Sample mean & 0.001 & 0.054 & 0.054 & 0.043 \\
 &  & Sample median & 0.001 & 0.044 & 0.044 & 0.035 \\
 &  & \(Z\)-estimator, \(\beta=1.05\) & 0.001 & 0.041 & 0.041 & 0.033 \\
 &  & \(Z\)-estimator, \(\beta=1.10\) & 0.001 & 0.042 & 0.042 & 0.033 \\
 &  & \(Z\)-estimator, \(\beta=1.25\) & 0.001 & 0.042 & 0.042 & 0.033 \\
 &  & \(Z\)-estimator, \(\beta=1.50\) & 0.001 & 0.043 & 0.043 & 0.034 \\
 &  & \(Z\)-estimator, \(\beta=1.75\) & 0.001 & 0.044 & 0.044 & 0.035 \\
 &  & \(Z\)-estimator, \(\beta=2.00\) & 0.001 & 0.044 & 0.044 & 0.035 \\
\addlinespace
5.0 & 1000 & Sample mean & 0.000 & 0.041 & 0.041 & 0.033 \\
 &  & Sample median & 0.000 & 0.042 & 0.042 & 0.034 \\
 &  & \(Z\)-estimator, \(\beta=1.05\) & 0.000 & 0.037 & 0.037 & 0.030 \\
 &  & \(Z\)-estimator, \(\beta=1.10\) & 0.000 & 0.037 & 0.037 & 0.030 \\
 &  & \(Z\)-estimator, \(\beta=1.25\) & 0.000 & 0.037 & 0.037 & 0.030 \\
 &  & \(Z\)-estimator, \(\beta=1.50\) & 0.000 & 0.037 & 0.037 & 0.030 \\
 &  & \(Z\)-estimator, \(\beta=1.75\) & 0.000 & 0.038 & 0.038 & 0.030 \\
 &  & \(Z\)-estimator, \(\beta=2.00\) & 0.000 & 0.038 & 0.038 & 0.030 \\
\bottomrule
\end{tabular}
\end{table}

\section{Real-data implementation details}
\label{app:realdata_details}

\subsection{freMTPL claim-severity processing}

The freMTPL data are provided in two linked files. The frequency file
\texttt{freMTPLfreq} contains one row per policy, including the policy identifier,
claim count, exposure, vehicle and driver characteristics, and geographical density.
The severity file \texttt{freMTPLsev} contains one row per individual claim payment and
the same policy identifier. We merge each payment record with the corresponding policy
covariates and use individual payments, rather than policy-level sums, as severity
observations. The full portfolio contains 413,169 policies, of which 96.28\% have no
recorded payment; the severity analysis therefore models only the 16,181 positive
payment records belonging to 15,390 policies with at least one claim.

The numerical fit includes an intercept and four policy covariates:
\texttt{Exposure}, \texttt{CarAge}, \texttt{DriverAge}, and \texttt{Density}. We
exclude \texttt{ClaimNb} because it is the realized frequency outcome rather than a
pre-specified severity rating covariate. The non-intercept covariates are standardized
using the training subset only. The response is the log claim amount,
\(\log(\texttt{ClaimAmount})\), the standard actuarial severity scale; the additive
two-component regression is fitted on this scale, and predictions are exponentiated back to
euros only for reporting. The categorical variables \texttt{Power}, \texttt{Brand}, \texttt{Gas},
and \texttt{Region} are omitted in this numerical implementation but can be incorporated
through dummy variables in the same likelihood.

Policies with observed payments are split into training, validation, and test subsets
using a 60/20/20 allocation with random seed 2026. Multiple claim records belonging to
the same policy are always assigned to the same subset, so policy covariates do not leak
across held-out subsets. To represent tail severity in each split, allocation is
stratified by deciles of the maximum \(\log(1+\texttt{ClaimAmount})\) within each
policy. The resulting split gives 9,234 training, 3,078 validation, and 3,078 test
policies, corresponding to 9,676, 3,258, and 3,247 individual payment observations,
respectively.

The finite-sample tail diagnostics in Figure~\ref{fig:fremtpl-claim-distribution}
should not be interpreted as a formal population tail classification. For any finite
sample, the empirical moment generating function
\(\widehat M_n(t)=n^{-1}\sum_i\exp(tY_i)\) is finite for every finite \(t\), so an
empirical MGF plot cannot establish the population condition
\(\mathbb{E}\{\exp(tY)\}=\infty\) for every \(t>0\).

\subsection{Binance return-innovation processing}
\label{app:binance_details}

The Binance experiment uses public Spot monthly 1-minute kline data from
\url{https://data.binance.vision/data/spot/monthly/klines/} for \texttt{BTCUSDT},
\texttt{ETHUSDT}, \texttt{BNBUSDT}, and \texttt{SOLUSDT}, from January 2021 through
December 2024. BTC, ETH, BNB, and SOL denote Bitcoin, Ether, the BNB token
(formerly Binance Coin), and Solana,
respectively, and USDT denotes the Tether quote currency. We stop before 2025 to avoid
mixing the millisecond timestamp files used
here with the newer microsecond timestamp convention. The 2021--2024 calendar contains
2,103,840 expected 1-minute timestamps. After merging the four symbols by open time,
2,102,766 timestamps are complete for all four close series; each symbol has about
1,073 missing 1-minute records over the four-year window. After constructing returns
and the lagged Bitcoin covariate, the 1-minute regression frame contains 2,102,748
observations.

\begin{figure}[htbp]
\centering
\includegraphics[width=0.85\textwidth]{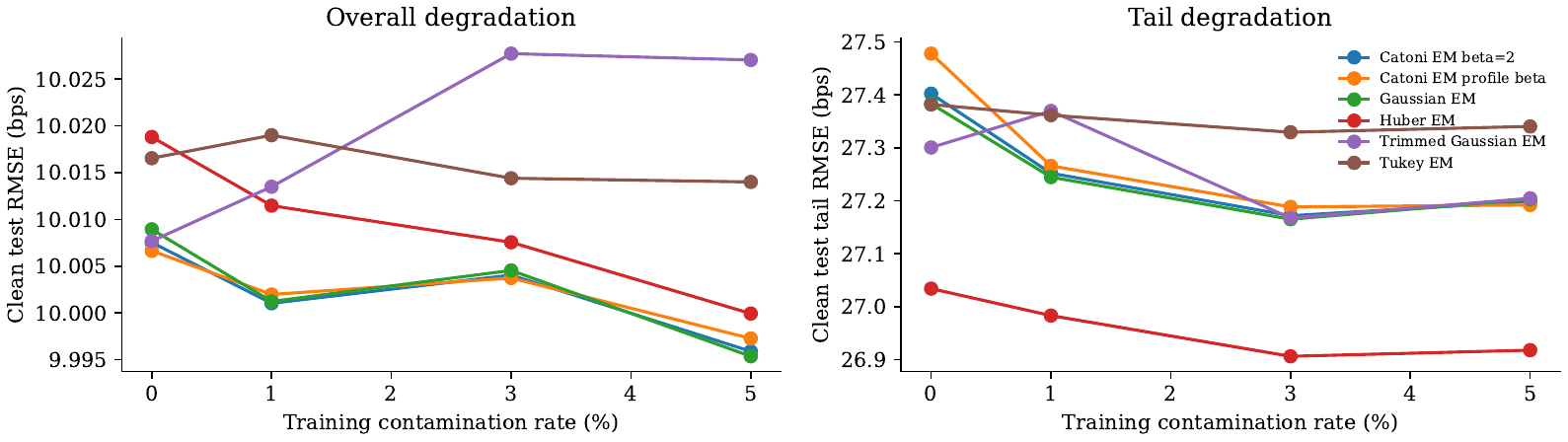}
\caption{Training-only symmetric spike contamination for the 5-minute Binance
experiment. The contaminated training responses are used only for fitting; validation
and test data remain clean.}
\label{fig:binance-crypto-contamination}
\end{figure}

We also aggregate closes to non-overlapping 5-minute intervals, giving 420,539
regression observations. All splits are chronological: 60\% training, 20\% validation,
and 20\% test. The 5-minute split has 252,323 training observations, 84,108 validation
observations, and 84,108 test observations. The clean 5-minute comparison in the main
text is fitted on all 252,323 training observations. For the larger 1-minute sensitivity
fit and for repeated training-contamination paths, the EM fits use 100,000
chronologically even training rows when the training split is larger, while validation
and test metrics are evaluated on the full chronological held-out splits.
The training-only contamination experiment perturbs only the fitting response in the
100,000-row training subset. We add symmetric return spikes of size 15 training standard
deviations, corresponding to 457.00 basis points in the 5-minute experiment. The clean
validation and test sets are unchanged. Figure~\ref{fig:binance-crypto-contamination}
shows the resulting clean-test degradation curves for contamination rates 0\%, 1\%,
3\%, and 5\%. Catoni and Gaussian EM remain close across the contamination path, Huber
gives the best tail RMSE, and the trimmed and Tukey fits are less stable in overall
RMSE at several contamination levels.

\section{Tail behavior and moment existence}
\label{app:tail_behavior}

\begin{proposition}[Tail behavior and moment existence]\label{prop:tail_behavior}
Let $X$ have density $f(x;\alpha,\beta)$ with $\alpha>0$ and $\beta\in(1,2]$. As
$z=|x|/\alpha\to\infty$,
\[
A_\beta(z)
=
\beta z\log z-(\beta+\log\beta)z+
\begin{cases}
O_\beta(z^{2-\beta}), & 1<\beta<2,\\
O(\log z), & \beta=2 .
\end{cases}
\]
Consequently,
\[
\log f(x;\alpha,\beta)
=
-\beta z\log z+(\beta+\log\beta)z+
\begin{cases}
O_\beta(z^{2-\beta}), & 1<\beta<2,\\
O(\log z), & \beta=2
\end{cases}
-\log(\alpha I_\beta),
\]
and, for every $\lambda>0$,
\[
\int_{\mathbb{R}} e^{\lambda |x|} f(x;\alpha,\beta)\,dx<\infty,
\]
and all polynomial moments of $X$ are finite. Thus \(f(\cdot;\alpha,\beta)\) is
super-exponentially light-tailed, not heavy-tailed under the standard moment-generating
function definition.
\end{proposition}

\begin{proof}
Let \(g_\beta(z)=\log(1+z+z^\beta/\beta)\) for \(z\geq0\).
For \(z\geq1\),
\[
g_\beta(z)
=
\log\left\{\frac{z^\beta}{\beta}
\left(1+\beta z^{1-\beta}+\beta z^{-\beta}\right)\right\}
=
\beta\log z-\log\beta+
\log\left(1+\beta z^{1-\beta}+\beta z^{-\beta}\right).
\]
Using \(0\leq\log(1+v)\leq v\) for \(v\geq0\), the integral of the final logarithm
from \(1\) to \(z\) is \(O_\beta(z^{2-\beta})\) for \(1<\beta<2\) and
\(O(\log z)\) for \(\beta=2\). Also,
\[
\int_1^z(\beta\log u-\log\beta)\,du
=
\beta z\log z-(\beta+\log\beta)z+O_\beta(1).
\]
Adding the finite integral over \([0,1]\) proves the expansion for \(A_\beta\), and
\(\log f(x;\alpha,\beta)=-A_\beta(|x|/\alpha)-\log(\alpha I_\beta)\) gives the
displayed log-density expansion.
For any fixed $\lambda>0$, the exponent in
$e^{\lambda |x|}f(x;\alpha,\beta)$ is eventually bounded above by
$-c|x|\log |x|$ for some $c>0$. Hence
\(\int_{\mathbb{R}} e^{\lambda |x|}f(x;\alpha,\beta)dx<\infty\).
All polynomial moments are finite because $|x|^k\leq C_{k,\lambda}e^{\lambda |x|}$ for
any $k>0$ and any $\lambda>0$.
\end{proof}

\section{Evaluation of the normalization constant}
\label{app:norm_const}

In this appendix, we provide the derivation of the normalization constant and describe
the numerical quadrature scheme used in the implementation.

Recall that
\(\psi_\beta(x)=\operatorname{sign}(x)\log(1+|x|+|x|^\beta/\beta)\) and
\(R_\beta(x)=2\int_0^x \psi_\beta(t)dt\).
Since $\psi_\beta$ is an odd function, $R_\beta$ is even. Indeed, for any $z>0$,
\(R_\beta(z)=2\int_0^z\log(1+u+u^\beta/\beta)du\),
whereas
\[
R_\beta(-z)
=
2\int_0^{-z}\psi_\beta(u)du
=
2\int_0^z
\log\left(1+u+\frac{u^\beta}{\beta}\right)du
=
R_\beta(z).
\]
Therefore, for all $z\in\mathbb{R}$,
\begin{equation}\label{eq:R_even_appendix}
R_\beta(z)
=
2\int_0^{|z|}
\log\left(1+u+\frac{u^\beta}{\beta}\right)du .
\end{equation}

Using the change of variables $z=x/\alpha$, we have
\[
C^{-1}_{\alpha,\beta}
=
\int_{-\infty}^{\infty}
\exp\left\{
-\frac{1}{2}R_\beta\left(\frac{x}{\alpha}\right)
\right\}dx
=
\alpha
\int_{-\infty}^{\infty}
\exp\left\{
-\frac{1}{2}R_\beta(z)
\right\}dz .
\]
By the evenness of $R_\beta$, this becomes
\[
C^{-1}_{\alpha,\beta}
=
2\alpha
\int_0^\infty
\exp\left\{
-\frac{1}{2}R_\beta(z)
\right\}dz .
\]
Combining this with \eqref{eq:R_even_appendix}, we obtain
\[
C^{-1}_{\alpha,\beta}
=
2\alpha
\int_0^\infty
\exp\left\{
-\int_0^z
\log\left(1+u+\frac{u^\beta}{\beta}\right)du
\right\}dz
=
\alpha I_\beta,
\]
where
$
I_\beta
=
2\int_0^\infty
\exp\left\{
-\int_0^z
\log\left(1+u+\frac{u^\beta}{\beta}\right)du
\right\}dz$. Thus,
\(C_{\alpha,\beta}=1/(\alpha I_\beta)\).

For numerical computation, define
\(g_\beta(u)=\log(1+u+u^\beta/\beta)\) and
\(A_\beta(z)=\int_0^z g_\beta(u)du\). Then
\(I_\beta=2\int_0^\infty \exp\{-A_\beta(z)\}dz\).
In practice, we truncate the outer integral at a sufficiently large value $L$ and compute
\(I_\beta(L)=2\int_0^L \exp\{-A_\beta(z)\}dz\).
Let $0=z_0<z_1<\cdots<z_M=L$ be an equally spaced grid with step size
$\Delta=L/M$. The inner integral $A_\beta(z)$ can be evaluated cumulatively by the
trapezoidal rule with \(\widehat A_\beta(z_0)=0\),
and, for $j=1,\ldots,M$,
\[
\widehat A_\beta(z_j)
=
\widehat A_\beta(z_{j-1})
+
\frac{\Delta}{2}
\left\{
g_\beta(z_{j-1})+g_\beta(z_j)
\right\}.
\]
The outer integral is then approximated by
\[
\widehat I_\beta
=
2\Delta
\left[
\frac{1}{2}\exp\{-\widehat A_\beta(z_0)\}
+
\sum_{j=1}^{M-1}\exp\{-\widehat A_\beta(z_j)\}
+
\frac{1}{2}\exp\{-\widehat A_\beta(z_M)\}
\right].
\]
Finally, the normalization constant is approximated by
\(\widehat C_{\alpha,\beta}=1/(\alpha \widehat I_\beta)\).

This cumulative implementation avoids repeatedly computing the inner integral from
scratch for each grid point. Hence, for a fixed $\beta$, $\widehat I_\beta$ can be
precomputed once and reused for all scale parameters $\alpha$.

\section{Derivative for the scale update}
\label{app:alpha_update}

This appendix derives the first-order condition used in the componentwise scale update.
For component $i$, define the residual
\(r_{ij}^{(t+1)}=Y_j-X_j^\top b_i^{(t+1)}\)
and consider the $\alpha_i$-dependent part of the EM objective,
\begin{equation}\label{eq:alpha_Q_appendix}
Q_i^{(t)}(\alpha_i)
=
\sum_{j=1}^n w_{ij}^{(t)}
\left[
\log C_{\alpha_i,\beta}
-
\frac{1}{2}
R_\beta\!\left(\frac{r_{ij}^{(t+1)}}{\alpha_i}\right)
\right],
\qquad \alpha_i>0 .
\end{equation}

By the location--scale representation of the normalization constant,
$C_{\alpha,\beta}=1/(\alpha I_\beta)$, where $I_\beta$ does not depend on $\alpha$.
Hence \(\frac{\partial}{\partial \alpha}\log C_{\alpha,\beta}=-1/\alpha\).
Moreover, since $R_\beta'(u)=2\psi_\beta(u)$, the chain rule gives
\[
\frac{\partial}{\partial \alpha}
R_\beta\!\left(\frac{x}{\alpha}\right)
=
-\frac{2x}{\alpha^2}
\psi_\beta\!\left(\frac{x}{\alpha}\right).
\]
Applying these two identities to \eqref{eq:alpha_Q_appendix}, we obtain
\begin{equation}\label{eq:alpha_derivative_appendix}
\frac{d}{d\alpha_i}Q_i^{(t)}(\alpha_i)
=
\sum_{j=1}^n w_{ij}^{(t)}
\left[
-\frac{1}{\alpha_i}
+
\frac{r_{ij}^{(t+1)}}{\alpha_i^2}
\psi_\beta\!\left(\frac{r_{ij}^{(t+1)}}{\alpha_i}\right)
\right].
\end{equation}
Therefore, the first-order condition is
\[
\sum_{j=1}^n w_{ij}^{(t)}
\left[
\frac{r_{ij}^{(t+1)}}{\alpha_i}
\psi_\beta\!\left(\frac{r_{ij}^{(t+1)}}{\alpha_i}\right)
-1
\right]
=0 .
\]
Using
\[
\frac{r}{\alpha_i}
\psi_\beta\!\left(\frac{r}{\alpha_i}\right)
=
\frac{|r|}{\alpha_i}
\log\!\left(
1+\frac{|r|}{\alpha_i}
+\frac{1}{\beta}
\left(\frac{|r|}{\alpha_i}\right)^\beta
\right),
\]
the first-order condition can be written as the scalar root equation
\begin{equation}\label{eq:alpha_root_appendix}
\sum_{j=1}^n w_{ij}^{(t)}
\left[
\frac{|r_{ij}^{(t+1)}|}{\alpha_i}
\log\!\left(
1+\frac{|r_{ij}^{(t+1)}|}{\alpha_i}
+\frac{1}{\beta}
\left(\frac{|r_{ij}^{(t+1)}|}{\alpha_i}\right)^\beta
\right)
-1
\right]
=0 .
\end{equation}
This is the root equation used in Theorem~\ref{thm:alpha_root_componentwise}.

\section{Derivative for the optional shape profile}
\label{app:beta_profile_derivative}

When the common shape parameter is estimated, a derivative formula provides a check on
the one-dimensional profile search. For \(u\geq0\), let
\[
g_\beta(u)=\log(1+u+u^\beta/\beta),
\qquad
A_\beta(z)=\int_0^z g_\beta(u)\,du .
\]
For \(u>0\),
\[
\partial_\beta g_\beta(u)
=
\frac{u^\beta(\beta\log u-1)}
{\beta^2(1+u+u^\beta/\beta)},
\qquad
\partial_\beta A_\beta(z)
=
\int_0^z
\frac{u^\beta(\beta\log u-1)}
{\beta^2(1+u+u^\beta/\beta)}\,du .
\]
Let \(S_\beta\) have density \(2e^{-A_\beta(s)}/I_\beta\) on
\([0,\infty)\). Differentiation under the integral sign gives
\[
\partial_\beta\log I_\beta
=
-\mathbb E\{\partial_\beta A_\beta(S_\beta)\}.
\]
Therefore, for fixed responsibilities, residuals, and scales, the common-shape score is
\[
\frac{\partial Q}{\partial\beta}
=
-\left(\sum_{i=1}^m\sum_{j=1}^n w_{ij}\right)
\partial_\beta\log I_\beta
-
\sum_{i=1}^m\sum_{j=1}^n w_{ij}
\partial_\beta A_\beta\left(\frac{|r_{ij}|}{\alpha_i}\right).
\]
When each \(\widehat\alpha_i(\beta)\) is an interior differentiable scale maximizer, the
envelope theorem shows that the same expression evaluated at
\(\widehat\alpha_i(\beta)\) is the derivative of the profiled objective
\eqref{eq:beta_profile_objective_main}. This formula does not remove boundary behavior:
a maximum at or near \(\beta_{\min}\) remains evidence of weak shape identification.

\section{Details of the rejection sampling scheme}
\label{app:sampling}

We next give a rejection sampler for the proposed density. The lower bound of
Lemma~\ref{upbound}, namely \(R_\beta(x/\alpha)\geq 2\phi(|x|/\alpha)\) with
\(\phi(s)=(1+s)\log(1+s)-s\), yields the envelope
\[
f(x;\alpha,\beta)
=
C_{\alpha,\beta}\exp\{-R_\beta(x/\alpha)/2\}
\leq
C_{\alpha,\beta}\exp\{-\phi(|x|/\alpha)\}
=
C_{\alpha,\beta}
\frac{\exp(|x|/\alpha)}
{(1+|x|/\alpha)^{1+|x|/\alpha}} .
\]
Let
\[
q(x;\alpha)
=
\frac{1}{D_\alpha}
\frac{\exp(|x|/\alpha)}
{(1+|x|/\alpha)^{1+|x|/\alpha}},
\]
where
$
D_\alpha
=
\int_{-\infty}^{\infty}
\frac{\exp(|x|/\alpha)}
{(1+|x|/\alpha)^{1+|x|/\alpha}}dx$. Then
\[
\frac{f(x;\alpha,\beta)}{q(x;\alpha)}
\leq
C_{\alpha,\beta}D_\alpha
=:M .
\]

Using $D_\alpha=\alpha D_0$ and $C_{\alpha,\beta}=1/(\alpha I_\beta)$, we have
\[
M=\frac{D_0}{I_\beta},
\qquad
D_0=
2\int_0^\infty
\frac{\exp(s)}
{(1+s)^{1+s}}ds .
\]
Thus, for a fixed $\beta$, the rejection constant can be computed once.

To sample from $q(x;\alpha)$, write $s=|x|/\alpha$. The density of $s$ is proportional to
\(\exp(s)/(1+s)^{1+s}
=\exp\{-[(1+s)\log(1+s)-s]\}\).
Recall \(\phi(s)=(1+s)\log(1+s)-s\) for \(s\geq 0\) from Lemma~\ref{upbound}.
Since $\phi''(s)=1/(1+s)>0$, the density proportional to $\exp\{-\phi(s)\}$ is
log-concave and can be sampled by adaptive rejection sampling.

\begin{algorithm}[H]
\caption{Rejection sampler for $f(\cdot;\alpha,\beta)$}
\label{alg:proposed_rejection_sampler}
\begin{algorithmic}[1]
\REQUIRE Scale parameter $\alpha>0$, shape parameter $\beta\in(1,2]$, precomputed constants
$I_\beta$ and $D_0$.
\ENSURE A random sample $X\sim f(\cdot;\alpha,\beta)$.
\STATE Set $M=D_0/I_\beta$.
\REPEAT
    \STATE Generate $S$ from the density on $[0,\infty)$ proportional to
    \[
    \exp\left\{-[(1+s)\log(1+s)-s]\right\}.
    \]
    \STATE Generate $B\in\{-1,1\}$ with $\Pr(B=1)=\Pr(B=-1)=1/2$.
    \STATE Set $Y=\alpha B S$.
    \STATE Generate $U\sim \mathrm{Unif}(0,1)$.
    \STATE Compute
    \[
    a(Y)
    =
    \exp\left\{
    -\frac{1}{2}R_\beta(Y/\alpha)
    +
    (1+|Y|/\alpha)\log(1+|Y|/\alpha)
    -
    |Y|/\alpha
    \right\}.
    \]
\UNTIL{$U\leq a(Y)$}
\STATE \RETURN $X=Y$.
\end{algorithmic}
\end{algorithm}

\section{Proof of Lemma \ref{upbound}}
\label{app:proof_upbound}
\begin{proof}
Since $R_\beta$ is even, it suffices to consider $x\ge 0$ and write
\(R_\beta(x/\alpha)=2\int_{0}^{x/\alpha}\log(1+t+t^\beta/\beta)\,dt\).
Because \(1+t+t^\beta/\beta\geq1+t\) for \(t\geq0\),
\[
R_\beta(x/\alpha)
\geq
2\int_0^{x/\alpha}\log(1+t)\,dt
=2\phi(x/\alpha).
\]
For $t\ge 0$ we use the submultiplicative bound
\(1+t+t^\beta/\beta \le (1+t)(1+t^\beta/\beta)\),
whence, by monotonicity of $\log$,
\(\log(1+t+t^\beta/\beta)\le\log(1+t)+\log(1+t^\beta/\beta)\).
Applying $\log(1+u)\le u$ for $u\ge 0$ gives
\(\log(1+t+t^\beta/\beta)\le t+t^\beta/\beta\).
Therefore,
\[
R_\beta(x/\alpha)
\;\le\;2\int_{0}^{x/\alpha}\!\!\Bigl(t+\frac{t^\beta}{\beta}\Bigr)\,dt
\;=\;\Bigl(\frac{x}{\alpha}\Bigr)^{\!2}
\;+\;\frac{2}{\beta(\beta+1)}\Bigl(\frac{x}{\alpha}\Bigr)^{\!\beta+1},
\]
which yields the stated bound (and symmetry replaces $x$ by $|x|$).
\end{proof}

\section{Proof of Proposition~\ref{prop:pi_lower_bound}}
\label{app:proof_pi_step}

\begin{proof}
At a fixed EM iteration, suppress the superscript $t$ and write $W_i=W_i^{(t)}$. The
mixing-weight block maximizes the strictly concave function
\[
\sum_{i=1}^m W_i\log\pi_i
\]
over the convex compact set $\{\pi_i\geq\pi_{\min},\sum_i\pi_i=1\}$. Hence every KKT
point is a global maximizer, and the maximizer is unique on the coordinates with
$W_i>0$.

Let $\lambda$ be the multiplier for $\sum_i\pi_i=1$ and let $\nu_i\ge0$ be the multiplier
for $\pi_i\geq\pi_{\min}$. The stationarity condition is
\[
\frac{W_i}{\pi_i}-\lambda+\nu_i=0,
\qquad
\nu_i(\pi_i-\pi_{\min})=0 .
\]
If the lower bound is inactive, then $\nu_i=0$ and $\pi_i=W_i/\lambda$. If it is active,
then $\pi_i=\pi_{\min}$ and the stationarity condition implies
$W_i/\pi_{\min}\leq\lambda$, equivalently $W_i/\lambda\leq\pi_{\min}$. Therefore the KKT
solution has the form
\[
\pi_i=\max\left\{\pi_{\min},\frac{W_i}{\lambda}\right\}.
\]
It remains only to choose $\lambda$ so that the simplex constraint holds.

The function
\[
F(\lambda)=\sum_{i=1}^m\max\left\{\pi_{\min},\frac{W_i}{\lambda}\right\}
\]
is continuous and nonincreasing for $\lambda>0$. Since $\sum_iW_i=n>0$,
$F(\lambda)\to\infty$ as $\lambda\downarrow0$ over the positive-weight components, while
$F(\lambda)\to m\pi_{\min}<1$ as $\lambda\to\infty$. Thus there is a solution to
$F(\lambda)=1$. The crossing is unique because, at any solution with
$1>m\pi_{\min}$, at least one component must be inactive, and on the inactive set
$F$ is strictly decreasing in $\lambda$. This proves \eqref{eq:pi_kkt_update}.
\end{proof}

\section{Proof of Theorem~\ref{thm:alpha_root_componentwise}}
\label{app:proof_alpha_root}

\begin{proof}
Let $r_j=r_{ij}^{(t+1)}$, $w_j=w_{ij}^{(t)}$, and
$W_i^{(t)}=\sum_{j=1}^n w_j$. From the calculation of the derivative in \ref{app:alpha_update}, the derivative of
the objective of the component scale satisfies
\[
\frac{d}{d\alpha_i}Q_i^{(t)}(\alpha_i)
=
\frac{1}{\alpha_i}
G_i^{(t)}(\alpha_i),
\]
where
\[
G_i^{(t)}(\alpha_i)
=
\sum_{j=1}^n
w_j h_\beta\left(\frac{|r_j|}{\alpha_i}\right)
-
W_i^{(t)},
\qquad
h_\beta(u)=u\log\left(1+u+\frac{u^\beta}{\beta}\right).
\]

It remains to analyze $G_i^{(t)}$. For $u>0$,
\[
h_\beta'(u)
=
\log\left(1+u+\frac{u^\beta}{\beta}\right)
+
u\frac{1+u^{\beta-1}}{1+u+u^\beta/\beta}
>0,
\]
so $h_\beta$ is strictly increasing on $[0,\infty)$. Since $|r_j|/\alpha_i$ is decreasing
in $\alpha_i$, if
\[
\sum_{j=1}^n w_j\mathbf{1}\{|r_j|>0\}>0,
\]
then $G_i^{(t)}$ is continuous and strictly decreasing on $(0,\infty)$.

Moreover, as $\alpha_i\to\infty$, $|r_j|/\alpha_i\to 0$ and hence
$G_i^{(t)}(\alpha_i)\to -W_i^{(t)}<0$. As $\alpha_i\to 0^+$, at least one term
$h_\beta(|r_j|/\alpha_i)$ diverges to $+\infty$, so
$G_i^{(t)}(\alpha_i)\to+\infty$. Therefore, by continuity and strict monotonicity,
there exists a unique $\widetilde\alpha_i\in(0,\infty)$ such that
$G_i^{(t)}(\widetilde\alpha_i)=0$.

Finally, since $\alpha_i>0$, the sign of $dQ_i^{(t)}(\alpha_i)/d\alpha_i$ is the sign of
$G_i^{(t)}(\alpha_i)$. Thus $Q_i^{(t)}$ increases before $\widetilde\alpha_i$ and decreases
after $\widetilde\alpha_i$, which proves that $\widetilde\alpha_i$ is the unique
unconstrained global maximizer. The constrained maximizer over
$[\alpha_{\min},\alpha_{\max}]$ is therefore the Euclidean clipping of
$\widetilde\alpha_i$ to this interval.

If all positively weighted residuals are zero, then
\[
Q_i^{(t)}(\alpha_i)
=
\sum_{j=1}^n w_j[-\log\alpha_i-\log I_\beta]
=
-W_i^{(t)}\log\alpha_i-W_i^{(t)}\log I_\beta .
\]
This function has no maximizer on $(0,\infty)$ because it increases as
$\alpha_i\downarrow0$. On the constrained interval
$[\alpha_{\min},\alpha_{\max}]$, it is strictly decreasing, so its maximizer is
$\alpha_{\min}$.
\end{proof}

\section{Proof of Theorem~\ref{EMmonotonically1}}
\label{app:proof_em_monotonicity}

\begin{proof}
Let
$
p_{ij}(\Theta)
=
\frac{\pi_i f(Y_j-X_j^\top b_i;\alpha_i,\beta)}
{\sum_{l=1}^m \pi_l f(Y_j-X_j^\top b_l;\alpha_l,\beta)}
$
be the posterior probability of component $i$ under parameter $\Theta$, and let
$q_{ij}=w_{ij}^{(t)}=p_{ij}(\Theta^{(t)})$. For any $\Theta$,
\[
\ell(Y,X\mid\Theta)
=
Q(\Theta\mid\Theta^{(t)})
+
\sum_{j=1}^n
\mathrm{KL}\big(q_j\,\|\,p_j(\Theta)\big),
\]
where $q_j=(q_{1j},\ldots,q_{mj})$ and $p_j(\Theta)=(p_{1j}(\Theta),\ldots,p_{mj}(\Theta))$.
At $\Theta=\Theta^{(t)}$, the KL term is zero, and hence
\[
\ell(Y,X\mid\Theta^{(t)})
=
Q(\Theta^{(t)}\mid\Theta^{(t)}).
\]
Therefore,
\[
\ell(Y,X\mid\Theta^{(t+1)})
-
\ell(Y,X\mid\Theta^{(t)})
=
Q(\Theta^{(t+1)}\mid\Theta^{(t)})
-
Q(\Theta^{(t)}\mid\Theta^{(t)})
+
\sum_{j=1}^n
\mathrm{KL}\big(q_j\,\|\,p_j(\Theta^{(t+1)})\big).
\]
Since the KL divergence is nonnegative, any update satisfying
\[
Q(\Theta^{(t+1)}\mid\Theta^{(t)})
\geq
Q(\Theta^{(t)}\mid\Theta^{(t)})
\]
also satisfies
$
\ell(Y,X\mid\Theta^{(t+1)})
\geq
\ell(Y,X\mid\Theta^{(t)})$.
\end{proof}

\section{Proofs for the population-to-sample EM guarantees}
\label{app:em_population_sample_guarantees}

\subsection{Proof of Theorem~\ref{thm:robust_population_contraction}}
\begin{proof}
The proof follows the perturbative argument for population EM maps in
\citet{balakrishnan2017statistical}, written in the notation of the present constrained
mixture regression. Since \(\theta^\star\) maximizes
\(q(\theta')=\bar Q(\theta'\mid\theta^\star)\) over the local convex chart, the
variational inequality gives
\begin{equation}\label{eq:vi_star_robust}
\left\langle
\nabla_{\theta'}\bar Q(\theta^\star\mid\theta^\star),
\theta'-\theta^\star
\right\rangle\le 0,
\qquad \theta'\in\mathcal C_r.
\end{equation}
Similarly, since \(M(\theta)\) maximizes
\(\theta'\mapsto \bar Q(\theta'\mid\theta)\) over the same chart,
\begin{equation}\label{eq:vi_M_robust}
\left\langle
\nabla_{\theta'}\bar Q(M(\theta)\mid\theta),
\theta'-M(\theta)
\right\rangle\le 0,
\qquad \theta'\in\mathcal C_r.
\end{equation}
Taking \(\theta'=M(\theta)\) in \eqref{eq:vi_star_robust} and
\(\theta'=\theta^\star\) in \eqref{eq:vi_M_robust}, and adding the two inequalities,
yields
\begin{align}
&\left\langle
\nabla_{\theta'}\bar Q(M(\theta)\mid\theta^\star)
-
\nabla_{\theta'}\bar Q(\theta^\star\mid\theta^\star),
\theta^\star-M(\theta)
\right\rangle \notag\\
&\qquad\le
\left\langle
\nabla_{\theta'}\bar Q(M(\theta)\mid\theta^\star)
-
\nabla_{\theta'}\bar Q(M(\theta)\mid\theta),
\theta^\star-M(\theta)
\right\rangle .\label{eq:basic_pop_robust}
\end{align}
By \(\lambda\)-strong concavity of \(q\),
\begin{equation}\label{eq:strong_concavity_inner_robust}
\left\langle
\nabla_{\theta'}\bar Q(M(\theta)\mid\theta^\star)
-
\nabla_{\theta'}\bar Q(\theta^\star\mid\theta^\star),
\theta^\star-M(\theta)
\right\rangle
\ge
\lambda\|M(\theta)-\theta^\star\|_2^2 .
\end{equation}
By the first-order stability condition \eqref{eq:FOS_robust} and Cauchy's inequality,
\begin{align}
&\left\langle
\nabla_{\theta'}\bar Q(M(\theta)\mid\theta^\star)
-
\nabla_{\theta'}\bar Q(M(\theta)\mid\theta),
\theta^\star-M(\theta)
\right\rangle \notag\\
&\qquad\le
\gamma\|\theta-\theta^\star\|_2\|M(\theta)-\theta^\star\|_2 .\label{eq:FOS_inner_robust}
\end{align}
Combining \eqref{eq:basic_pop_robust}, \eqref{eq:strong_concavity_inner_robust}, and
\eqref{eq:FOS_inner_robust}, then cancelling the common factor
\(\|M(\theta)-\theta^\star\|_2\) when it is nonzero, gives
\[
\|M(\theta)-\theta^\star\|_2
\le
\frac{\gamma}{\lambda}\|\theta-\theta^\star\|_2 .
\]
If \(M(\theta)=\theta^\star\), the same inequality is immediate. The iterated population
bound follows by induction.
\end{proof}

\subsection{Proof of Theorem~\ref{thm:robust_sample_em}}
\begin{proof}
Work on the event
\[
\mathcal E_n
=
\left\{
\sup_{\theta\in\mathcal C_r}
\|M_n(\theta)-M(\theta)\|_2
\le
\epsilon_M^{\mathrm{unif}}(n,\delta)
\right\},
\]
which has probability at least \(1-\delta\) by the definition of
\(\epsilon_M^{\mathrm{unif}}(n,\delta)\). On this event, for every
\(\theta\in\mathcal C_r\),
\begin{align}
\|\widetilde M_n(\theta)-\theta^\star\|_2
&\le
\|M(\theta)-\theta^\star\|_2
+
\|M_n(\theta)-M(\theta)\|_2
+
\|\widetilde M_n(\theta)-M_n(\theta)\|_2 \notag\\
&\le
\kappa\|\theta-\theta^\star\|_2
+
\epsilon_M^{\mathrm{unif}}(n,\delta)+a_n .\label{eq:one_step_sample_robust}
\end{align}
Because \(\theta^0\in\mathcal C_r\) and
\(\epsilon_M^{\mathrm{unif}}(n,\delta)+a_n\le(1-\kappa)r\), the bound
\eqref{eq:one_step_sample_robust} implies
\[
\|\theta^1-\theta^\star\|_2
\le
\kappa r+(1-\kappa)r=r .
\]
Thus \(\theta^1\) remains in the local radius; because the implemented update is
constrained to the same local chart, \(\theta^1\in\mathcal C_r\). Repeating the same
argument by induction, all iterates remain in \(\mathcal C_r\) and satisfy
\[
\|\theta^{s+1}-\theta^\star\|_2
\le
\kappa\|\theta^s-\theta^\star\|_2
+
\epsilon_M^{\mathrm{unif}}(n,\delta)+a_n .
\]
Iterating this scalar recursion gives
\begin{align*}
\|\theta^t-\theta^\star\|_2
&\le
\kappa^t\|\theta^0-\theta^\star\|_2
+
\sum_{\ell=0}^{t-1}\kappa^\ell
\{\epsilon_M^{\mathrm{unif}}(n,\delta)+a_n\}\\
&\le
\kappa^t\|\theta^0-\theta^\star\|_2
+
\frac{\epsilon_M^{\mathrm{unif}}(n,\delta)+a_n}{1-\kappa},
\end{align*}
which proves \eqref{eq:sample_bound_robust}. The terminal bound
\eqref{eq:sample_terminal_robust} follows by choosing \(T\) so that the geometric term is
no larger than the second term.
\end{proof}

\subsection{Proof of Theorem~\ref{thm:robust_rate}}
\begin{proof}
The proof gives the details behind the \(n^{-1/2}\)-type local statistical radius. For
\(\theta,\theta'\in\mathcal C_r\), define the score field
\[
S(O;\theta',\theta)
=
\nabla_{\theta'}
\sum_{i=1}^{m}
w_i(O;\theta)
\Bigl\{
\log\pi_i'
+
\log f(Y-X^\top b_i';\exp(\eta_i'),\beta)
\Bigr\}.
\]
The proposed density satisfies
\[
\partial_{b_i}\log f(Y-X^\top b_i;e^{\eta_i},\beta)
=
\frac{X}{e^{\eta_i}}
\psi_\beta\left(\frac{Y-X^\top b_i}{e^{\eta_i}}\right),
\]
and
\[
\partial_{\eta_i}\log f(Y-X^\top b_i;e^{\eta_i},\beta)
=
-1+
\frac{Y-X^\top b_i}{e^{\eta_i}}
\psi_\beta\left(\frac{Y-X^\top b_i}{e^{\eta_i}}\right).
\]
Since
\[
|\psi_\beta(u)|
\le
C_\beta\log(2+|u|),
\qquad
0\le \psi_\beta'(u)\le C_\beta,
\]
and the local chart keeps \(\pi_i\) away from zero and
\(\alpha_i=e^{\eta_i}\) away from zero and infinity, every coordinate of
\(S(O;\theta',\theta)\) is dominated by a constant multiple of
\[
1+\|X\|_2^2+|Y|+\|X\|_2\log(2+|Y|+\|X\|_2).
\]
The same type of envelope controls the Lipschitz constant of
\(S(O;\theta',\theta)\) in \((\theta',\theta)\). Under the stated sub-Gaussian design and
correct specification, \(Y=X^\top b_Z^\star+\epsilon_Z\), where \(Z\) is the latent
component label, is sub-exponential in the Orlicz sense; the proposed error density has
finite sub-exponential envelopes by the
tail result in Proposition~\ref{prop:tail_behavior}. Hence there is a constant \(K\)
such that for every coordinate \(k\),
\begin{equation}\label{eq:score_orlicz_bound}
\|S_k(O;\theta',\theta)\|_{\psi_1}\le K,
\qquad
\|S_k(O;\theta'_1,\theta_1)-S_k(O;\theta'_2,\theta_2)\|_{\psi_1}
\le
K\|(\theta'_1,\theta_1)-(\theta'_2,\theta_2)\|_2 .
\end{equation}
where \(\|U\|_{\psi_1}=\inf\{c>0:\mathbb E\exp(|U|/c)\le2\}\).

Let \(\mathcal N\) be a Euclidean \(n^{-2}\)-net of
$\mathbb{B}(r;\theta^\star)\times\mathbb{B}(r;\theta^\star)$ with cardinality at most
\((Cn)^{2d_\theta}\). Bernstein's inequality for sub-exponential variables, applied to
each coordinate and each point of the net, followed by a union bound, yields with
probability at least \(1-\delta\),
\begin{equation}\label{eq:net_gradient_bound}
\max_{(\theta',\theta)\in\mathcal N}
\left\|
\frac1n\sum_{j=1}^{n}S(O_j;\theta',\theta)
-
\mathbb E S(O;\theta',\theta)
\right\|_2
\le
C K
\sqrt{\frac{d_\theta\log(Cn)+\log(1/\delta)}{n}} .
\end{equation}
The Lipschitz part of \eqref{eq:score_orlicz_bound} extends
\eqref{eq:net_gradient_bound} from the net to the full product chart, changing only the
constant \(C\). Therefore, with the same probability,
\begin{equation}\label{eq:uniform_gradient_bound_robust}
\sup_{\theta,\theta'\in\mathcal C_r}
\left\|
\nabla_{\theta'}\bar Q_n(\theta'\mid\theta)
-
\nabla_{\theta'}\bar Q(\theta'\mid\theta)
\right\|_2
\le
C
\sqrt{\frac{d_\theta\log(Cn)+\log(1/\delta)}{n}} .
\end{equation}

It remains to convert the uniform gradient deviation into an operator deviation. Fix
\(\theta\in\mathcal C_r\) and write \(u=M_n(\theta)\) and \(v=M(\theta)\)
for the local sample and population maximizers. By the variational inequalities for the
two concave maximization problems,
\[
\langle\nabla_{\theta'}\bar Q(v\mid\theta),u-v\rangle\le 0,
\qquad
\langle\nabla_{\theta'}\bar Q_n(u\mid\theta),v-u\rangle\le 0 .
\]
The uniform lower Hessian bound gives strong monotonicity of the population gradient:
\[
\lambda_0\|u-v\|_2^2
\le
\left\langle
\nabla_{\theta'}\bar Q(v\mid\theta)-\nabla_{\theta'}\bar Q(u\mid\theta),
 u-v
\right\rangle .
\]
Combining the last three displays, we obtain
\begin{align*}
\lambda_0\|u-v\|_2^2
&\le
-\left\langle\nabla_{\theta'}\bar Q(u\mid\theta),u-v\right\rangle\\
&\le
\left\langle
\nabla_{\theta'}\bar Q_n(u\mid\theta)-\nabla_{\theta'}\bar Q(u\mid\theta),
 u-v
\right\rangle\\
&\le
\sup_{\theta,\theta'\in\mathcal C_r}
\left\|
\nabla_{\theta'}\bar Q_n(\theta'\mid\theta)
-
\nabla_{\theta'}\bar Q(\theta'\mid\theta)
\right\|_2
\|u-v\|_2 .
\end{align*}
After cancellation, \eqref{eq:uniform_gradient_bound_robust} implies
\eqref{eq:epsilon_rate_robust}. Substitution into Theorem~\ref{thm:robust_sample_em}
proves \eqref{eq:final_rate_robust}.
\end{proof}

\clearpage
\bibliographystyle{plainnat}
\bibliography{reference}

\end{document}